\documentclass[a4paper]{IEEEtran}
\usepackage{etex}
\usepackage{multirow, tabularx}
\usepackage{cite}
\usepackage[T1]{fontenc}
\usepackage{textcomp}
   \usepackage{graphicx}
   \usepackage{epstopdf}
  \graphicspath{{eps/}{png/}{jpg/}}
\usepackage[caption=false,font=footnotesize]{subfig}
\ifCLASSOPTIONcompsoc
\usepackage[caption=false,font=normalsize,labelfont=sf,textfont=sf]{subfig}
\else
\usepackage[caption=false,font=footnotesize]{subfig}
\fi
\usepackage{capt-of}   

\usepackage[utf8]{inputenc}
\usepackage[cmex10]{amsmath} 
\usepackage{array}
\usepackage{mdwmath}
\usepackage{bbm}
\usepackage{mdwtab}
\usepackage{eqparbox}
\usepackage{stfloats}
\usepackage{url}
\newtheorem{theorem}{Theorem}

\usepackage[noend]{algorithmic} 
\usepackage{algorithm,caption}
\usepackage{tikz}
\usepackage{pgfplots}
\tikzstyle{RectObject}=[rectangle,fill=white,draw,line width=0.5mm]
\tikzstyle{line}=[draw]
\tikzstyle{arrow}=[draw, -latex]
\usetikzlibrary{shapes,arrows}
\usetikzlibrary{fit,calc,positioning,decorations.pathreplacing,matrix}
\usepackage{verbatimbox}
\usepackage{boxhandler}
\usepackage{amssymb}
\usepackage{amsmath}
\usepackage{lipsum}
\usepackage{enumerate}
\def\A{\mathcal{A}}
\def\B{\mathcal{B}}
\def\U{\mathcal{U}}
\def\G{\mathcal{G}}
\def\M{\mathcal{M}}

\DeclareMathAlphabet\mathbfcal{OMS}{cmsy}{b}{n}
\def\X{\mathbfcal{X}}%
\def\Y{\mathbfcal{Y}}%
\def\Z{\mathbfcal{Z}}%
\def\T{\mathcal{T}}
\def\Y{\mathbfcal{Y}}
\def\M{\mathcal{M}}
\def\S{\mathcal{S}}
\def\R{\mathcal{R}}

\def\I{\mathcal{I}}

\newcommand{\Expect}{{\rm I\kern-.3em E}}
\title{Utility Fair Rate Allocation in  LTE/802.11 Networks}
\author{Bahar Partov, Douglas J. Leith\thanks{Work supported by SFI grants 11/PI/1177 and 13/RC/2077 and by Bell Labs Ireland.}\\Trinity College Dublin, Ireland }

\begin{document}
\maketitle
\begin{abstract}
We consider proportional fair rate allocation in a heterogeneous network with a mix of LTE and 802.11 cells which supports multipath and multihomed operation (simultaneous connection of a user device to multiple LTE BSs and 802.11 APs).  We show that the utility fair optimisation problem is non-convex but that a global optimum can be found by solving a sequence of convex optimisations in a distributed fashion.  The result is a principled approach to offload from LTE to 802.11 and for exploiting LTE/802.11 path diversity to meet user traffic demands.
\end{abstract}

\section{Introduction}
In this paper we consider rate allocation in a heterogeneous network with a mix of LTE and 802.11 cells.  Integrated design of LTE and 802.11 is topical in view of the continuing increases in data traffic, the fact that many cellular network operators also operate a large network of 802.11 hotspots and that user handsets are now typically equipped with both LTE and 802.11 interfaces.  Rather than an either/or handover-like setting where the question is which network to use our interest is instead in settings where user devices jointly use the LTE and 802.11 networks and may send data across both simultaneously.   Further, we consider situations where user devices may connect to multiple LTE Base Stations (BSs) and 802.11 Access Points (APs) simultaneously.  This allows us to encompass the Coordinated Multi Point (CoMP) features in release 11 of  LTE \cite{3GPPTR36.819} which allow coordinated transmission and reception across multiple BSs and also multihoming to multiple APs and through a single WLAN card as proposed in, for example, \cite{giustiniano2009clubadsl}.   Of course connection to a single BS or AP remains as a special case within this more general framework.   Simultaneous transmission across multiple interfaces might be implemented by routing each connection across one or other network (in a manner akin to load balancing) or by striping connections across both networks (e.g. via use of transport layer protocols such as multi-path TCP \cite{draft-ietf-mptcp-rfc6824bis}).

Our focus in this paper is on how to allocate the available LTE and 802.11 bandwidth amongst user devices in a heterogeneous network, and in particular how to determine a proportional fair rate allocation.  Our main contributions are as follows.  Firstly, we develop a throughput model for heterogeneous networks that include both LTE and 802.11 links and which encompasses multipath and multihomed operation (simultaneous connection of a user device to multiple LTE BSs and 802.11 APs).  We show that the rate region is non-convex, and is also not log-convex.  Our second contribution is therefore a sequential convex optimisation approach, based on determining a sequence of maximal convex subsets, that we show converges to a global optimum.   This optimisation approach is suited to distributed implementation.  Thirdly, we present a number of application examples demonstrating how this framework can be used to develop principled approaches to offload from LTE to 802.11 and for exploiting LTE/802.11 path diversity to meet user traffic demands.
\subsection{Motivating Example}
\begin{figure}[!htb]
\centering
\includegraphics[scale=1]{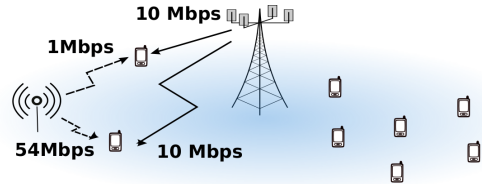}
\caption{Example of a mixed LTE/WiFi network, with one 802.11 AP, one LTE BS and two UEs $u_1$ and $u_2$ each equipped with both an 802.11 and an LTE interface. }
\label{fig:intro_example}
\end{figure}

Consider the simple example in Fig \ref{fig:intro_example}, where the network consists of one cellular BS and one  802.11 AP. UE $u_2$ is located close to the 802.11 AP and so uses a physical rate of 54Mbps. UE $u_1$ is further from the AP and so uses a lower physical rate of 1Mbps. Both UEs are located a similar distance from the LTE base station and use the same physical rate of 10Mbps for their LTE connection. The  physical rates available on the 802.11 and LTE links are summarised in Table \ref{tb:rates}.  For simplicity, we assume both UEs are saturated \emph{i.e.} always have a packet to send.

When the UEs are contending for the available resources, the achievable throughput is determined by the channel access method. 
For 802.11 the random access approach leads to collisions and reduced throughput when more than one station contends for access.  When $u_1$ and $u_2$ use only the 802.11 WLAN, their data rates are, respectively 0.46Mbps and 24.9Mbps for the standard 802.11n MAC settings (recall the UE physical rates are 1Mbps and 54Mbps). In the case of LTE, time-frequency slots are allocated to UEs by the cell BS.  When the UEs use only the LTE cell, the BS allocates a rate of 5 Mbps to each UE.
%
%

It can be seen that  $u_1$ enjoys better throughput via the LTE link than over 802.11, while $u_2$'s throughput over the 802.11 link is significantly better than that over the LTE connection. Due to this connection diversity the potential exists for more efficient decision-making  to improve performance. For this example we later show that if $u_1$ sends all traffic via  the BS, and $u_2$ sends all traffic via the AP, rates of 10 Mbps and 45 Mbps can be achieved by the UEs and that this is the proportional fair rate allocation. This is summarised in Table \ref{tb:rates}. Note that simply splitting the traffic for each UE equally over the 802.11 and LTE networks would yield rates of 5.46Mbps and 29.9Mbps \emph{i.e.} worse for \emph{both} users than the optimised allocation.
\begin{table}[!htb]
\centering
\caption{Motivating Example: Achievable data rates}
\label{tb:rates}
\begin{center}
\scalebox{0.9}{
\begin{tabular}{|c|c|c|c|c|c|}
\hline
\multirow{2}{.1cm}{}& \multicolumn{2}{p{2cm}|}{\centering PHY Rates [Mbps]} & \multicolumn{3}{p{2.8cm}|}{\centering Technology Rates [Mbps]} \\
\cline{2-6} & \multicolumn{1}{c|}{BS}  & \multicolumn{1}{c|}{AP}  & \multicolumn{1}{c|}{LTE Only} & \multicolumn{1}{c|}{802.11 Only} & \multicolumn{1}{c|}{Optimised Multi-RAT} \\ \hline
$u_1$  & 10 & 1 & 5 & 0.46 & 10 \\
$u_2$  & 10 & 54 & 5 & 24.9 & 42.4 \\ 
\hline
\end{tabular}
}
\end{center}
\end{table}
\section{Related Work}
The problem of resource allocation in heterogeneous wireless networks (HetNets) has been the subject of significant interest in the literature in the recent years.  
Much of this work has focussed on network selection, namely the task of selecting the best network for a UE to use.  See \cite{wang2013mathematical} for a survey on network selection in HetNets.   In network selection two modes of UE operation are typically considered: multi-modal and multi-homed. Although both multi-mode and multi-home operations support multiple radio interfaces, only the latter supports multiple TCP flows across disjoint paths. The task of network selection is usually based on a specified utility function with various methods (combinatorial, fuzzy logic, MDP, game theory, \emph{etc.})  proposed to solve the resulting utility-based network selection problem.
%
For example, \cite{blau2009decentralized}, \cite{ismail2012distributed} consider combinatorial optimisation, \cite{blau2009decentralized} considers a linear utility function in a multi-mode UMTS/GSM network and \cite{ismail2012distributed}  a linear utility function which is the sum of logarithmic utilities of individual BS-UE pairs.  In \cite{DBLP:journals/corr/abs-cs-0608025} a network-user association problem in a WLAN/UMTS hybrid network for multi-mode operations is solved using Markov Decision Processes (MDPs), although the complexity scales poorly with network size \cite{andrews2013overview}.   In \cite{aryafar2013rat} a non-cooperative game is formed among users of two different classes corresponding  to 3G/LTE and 802.11.
 
 Our work differs from this previous work in that (i) we consider concurrent association of the users  among OFDMA LTE BSs and 802.11 CSMA APs by considering both access network models, (ii) we formulate the problem as a proportional fair resource allocation problem, (iii) we show that although the problem is non-convex it can be solved by solving a sequence of convex optimisation methods and (iv) we use a low-complexity distributed optimisation method to solve each convex optimisation problem.
 
\section{Network model}
\subsection{Network topology}
We consider a fairly general multi-RAT setup where UEs can potentially connect to multiple LTE base stations and multiple 802.11 APs simultaneously.   Networks where UEs are constrained to connect to a single LTE BS and/or 802.11 AP are then special cases within this general setup.   We note that the Coordinated Multi-Point (CoMP) features in release 11 of LTE already allow transmission and reception across multiple BSs \cite{3GPPTR36.819},\cite{brueck2010centralized}.  
For 802.11 devices, connection to multiple APs might be implemented using the AP-hopping approach described in \cite{giustiniano2009clubadsl}, where an 802.11 client periodically hops between a set of APs, informing those APs with which it is not currently transmitting/receiving that it is in power-save mode so that these APs buffer packets directed to the client until it next connects. This software implementation is a client-side only change that does not require more than a single 802.11 radio and which uses standard hardware.

Let $\A$ denote the set of 802.11 access points {and $\U:=\{1,2,\dots, U\}$ the set of users.}   Associated with each access point $a\in \A$ is the set of users $\U_a\subset\U$ which can associate to it (so capturing geographic and access control constraints).   We also let $\A_u:=\{a\in\A: u\in\U_a\}\cup\{\emptyset\}$ denote the set of access points with which user $u$ can associate, where $\emptyset$ corresponds to the case where the user is not associated with any access point.   This also defines a bipartite graph $\G$ with one set of vertices corresponding to users, a second set of vertices corresponding to the access points and edges between each client vertex and the set of access point vertices to which it can connect.

Similarly, let $\B$ denote the set of LTE base stations, $\U_b$ the set of users located in the cell operated by $b\in\B$ and $\B_u$ the set of base stations with which user $u$ can connect. 


\subsection{LTE throughput}
Let $\I$ denote the set of available LTE sub-channels
and let $\zeta^i_{b,u}$ denote fraction of time sub-channel $i$ of BS $b$ is used by user $u$.   Let $\mathcal{E}^i$ be the set of user-BS pairs for which transmissions interfere on sub-channel $i$ (this defines a conflict graph).   We consider joint sub channel allocation and time sharing so that at a given time interfering BSs do not transmit on the same sub-channel \cite{huang2009scheduling}. Hence, the total allocation for each sub-channel must satisfy the following:
\begin{align}
\sum_{(u,b) \in \mathcal{E}^i} \zeta^{i}_{b,u} \leq 1 \quad  , \forall i \in \I \label{eq:c5}
\end{align}
When there is no frequency reuse, this constraint simplifies to $\sum_{u \in \U} \sum_{b\in \B}\zeta^i_{b,u} \leq 1,~\forall i \in \I$. Letting $\omega^i$ denote the sub-channel bandwidth, then the achievable rate of user $u$ is given by:
\begin{align}
r_{u}=\sum_{b \in \B} \sum_{i \in \I} \zeta^{i}_{b,u} \beta_1 \omega^i \log{(1+\frac{\gamma^{i}_{b,u}}{\beta_2})} \quad \forall u \in \U  \label{eq:c6}
\end{align}
with $\gamma^{i}_{b,u}=\frac{p^i_b e^{i}_{b,u}}{\sigma_n^2}$ denoting the SNR on sub-channel $i$ of BS $b$ where $p^i_b$,  $e^{i}_{b,u}$, $\sigma_n$ denote the BS power on channel $i$, channel gain, and noise power at the receiver respectively. {Parameters $\beta_1$ and $\beta_2$ reflect the LTE bandwidth and SINR implementation efficiencies \cite{mogensen2007lte}.}  Hence the rate region of the LTE network is the following 
\begin{align}
\R_{lte}= \bigg\{ \mathbf{r}:&  r_u= \sum_{b \in \B} \sum_{i \in \I} \zeta^{i}_{b,u} \beta_1 \omega^i \log{(1+\frac{\gamma^i_{b,u}}{ \beta_2})}, \nonumber \\
& \underline{r} \leq r_u \leq \bar{r},~0 \leq \zeta^{i}_{b,u} \leq 1,\notag\\
& \sum_{u \in \U} \sum_{b \in \B} \zeta^{i}_{b,u} \leq 1,~ \forall i \in \I \bigg\}
\end{align}
where $\mathbf{r}\in\mathbb{R}^U$ is the vector formed by stacking the user throughputs $r_u$, $u\in\U$ and we have also added the constraint that each user has maximum and minimum rates $\bar{r}$ and $\underline{r}$ respectively. Note that $\underline{r}$ may be $0$ and $\bar{r}$ may be $\infty$. It can be seen that the LTE rate region is convex (a polytope in fact, since the constraints are linear) in the time-frequency sharing factors $\zeta^i_{b,u}$. 

We note that this model can be extended to include dynamic power allocation per sub-channel, for example using a similar approach to  \cite{huang2009scheduling} and will still be convex.

\subsection{802.11 WLAN scheduling}\label{subsec:wlansch}
We consider an AP hopping approach where time is partitioned into scheduling slots of duration $T$ indexed by $t=1,2,\dots$ such that in slot $t$ user $u\in \U$ operates its 802.11 MAC in association with at most one access point (reflecting the constraint that user $u$ only has a single 802.11 radio).   Extension to allow simultaneous connection to multiple access points is straightforward. 

The access point which user $u$ selects at time slot $t$ is a random variable $\mathbf{A}_{u,t}$ which takes value $a$ from $\A_u\setminus\{\emptyset\}$ with probability $z_{a,u}$ (so $0\le z_{a,u}\le 1$ and $\sum_{a\in\A_u}z_{a,u}\le 1$).  Access points are selected independently at each slot and by each user.  Note that with this randomised schedule a client is not associated with any access point in a time slot (i.e. $\mathbf{A}_{u,t}=\emptyset$) with probability $p_{\emptyset,u}:=1-\sum_{a\in A_u}z_{a,u}$. 

\begin{figure}[h]
\centering
\begin{tikzpicture}

\node[draw=none] at (-5,0.5) {\small $t=1$};
\node[draw=none] at (-3.5,0.5) {\small $t=2$};
\node[draw=none] at (-2,0.5) {\small $t=3$};

\node[draw=none] at (-6.7,-.1) {\small $\mathbf{A}_{1,t}$};
\node (t1) at (-5,0) [draw,thick,minimum width=1.7cm,minimum height=.4cm] {\small $a_1$};
\node (t2) at (-3.3,0) [draw,thick,minimum width=1.7cm,minimum height=.4cm] {\small $a_2$};
\node (t3) at (-1.6,0) [draw,thick,minimum width=1.7cm,minimum height=.4cm] {\small $a_2$};

\node[draw=none] at (-6.7,-0.7) {\small $\mathbf{A}_{2,t}$};
\node (t4) at (-5,-0.6) [draw,thick,minimum width=1.7cm,minimum height=.4cm] {\small $a_1$};
\node (t5) at (-3.3,-0.6) [draw,thick,minimum width=1.7cm,minimum height=.4cm] {\small $a_1$};
\node (t6) at (-1.6,-0.6) [draw,thick,minimum width=1.7cm,minimum height=.4cm] {\small $a_2$};

\node[draw=none] at (-6.7,-1.3) {\small $\mathbf{N}_{1,t}$};
\node (t4) at (-5,-1.2) [draw,thick,minimum width=1.7cm,minimum height=.4cm] {\small $2$};
\node (t5) at (-3.3,-1.2) [draw,thick,minimum width=1.7cm,minimum height=.4cm] {\small $1$};
\node (t6) at (-1.6,-1.2) [draw,thick,minimum width=1.7cm,minimum height=.4cm] {\small $0$};

\node[draw=none] at (-6.7,-1.9) {\small $\mathbf{N}_{2,t}$};
\node (t4) at (-5,-1.8) [draw,thick,minimum width=1.7cm,minimum height=.4cm] {\small $0$};
\node (t5) at (-3.3,-1.8) [draw,thick,minimum width=1.7cm,minimum height=.4cm] {\small $1$};
\node (t6) at (-1.6,-1.8) [draw,thick,minimum width=1.7cm,minimum height=.4cm] {\small $2$};

\end{tikzpicture}
\caption{Illustrating the 802.11 random variables $\mathbf{A}_{u,t}$ and $\mathbf{N}_{a,t}$ related to user--AP association, for a two user, two AP example.}
\label{fig:80211_variables}
\end{figure}

Let random variable $\mathbf{N}_{a,t}$ denote the number of clients associated with access point $a$ in time slot $t$. {This is illustrated in  Fig \ref{fig:80211_variables} for a setup with two UEs, and two APs $a_1$, $a_2$. In slot $1$ both UEs are associated with AP $a_1$. In slot $2$ user $u_1$ is associated with AP $a_2$, while user $u_2$ is associated with AP $a_1$ and in slot $3$ both users are associated with AP $a_2$.} 

By construction,  random variables $\mathbf{N}_{a,t}$, $t=1,2,\dots$ are i.i.d. Let $2^{\S}$ denote the set consisting of the subsets of set $\S$, and let $\mathcal{P}_\kappa(\S)=\{s: s\in 2^{\S}, |s|=\kappa\}$ denote the subsets of set $\S$ which have cardinality $\kappa$. Therefore $\mathcal{P}_{n-1}(\U_a \setminus \{u\})$  is all subsets of $\U_a \setminus \{u\}$ with cardinality $n-1$. Define 
\begin{align}
p_{a,u,n}:=Prob(\mathbf{N}_{a,t}=n|\mathbf{A}_{u,t}=a)
\end{align}
to be the probability of $n$ users being associated with AP $a$ conditioned on user $u$ being associated with $a$. We have 

\begin{small}
\begin{align}
p_{a,u,n}=\begin{cases}
\prod\limits_{v\in \U_a\setminus\{u\}} (1-z_{a,v}) & n=1\\
 \sum\limits_{\stackrel{\tilde{\U}_a\in}{\mathcal{P}_{n-1}}(\U_a \setminus \{u\} )} \prod\limits_{v\in \tilde{\U}_a}z_{a,v}\prod\limits_{v\in \U_a\setminus\{\tilde{\U}_a,u\}} (1-z_{a,v}) & n>1
 \end{cases}
 \end{align}
 \end{small}
Letting $w_{u,a}=\frac{z_{a,u}}{1-z_{a,u}}$, this can be rewritten equivalently as
 \begin{align}
 p_{a,u,1} 
 &=
  \frac{1}{\prod\limits_{v\in \U_a\setminus\{u\}} (1+w_{a,v})}
 \end{align}
 and for $n>1$
 \begin{align}
 p_{a,u,n}
& =\sum\limits_{\stackrel{\tilde{\U}_a\in}{\mathcal{P}_{n-1}(\U_a \setminus \{u\} )}}  \prod\limits_{v\in \tilde{\U}_a}\frac{z_{a,v}}{1-z_{a,v}}\prod\limits_{v\in \U_a\setminus\{u\}} (1-z_{a,v})\\
& = \frac{\sum\limits_{\tilde{\U}_a\in\mathcal{P}_{n-1}(\U_a \setminus \{u\} )}  \prod\limits_{v\in \tilde{\U}_a}w_{a,v}}{\prod\limits_{v\in \U_a\setminus\{u\}} (1+w_{a,v})}
\end{align}
Defining,
\begin{align}
q_{a,u,n}:= 
\begin{cases}
1 & n=1\\
\sum\limits_{\tilde{\U}_a\in\mathcal{P}_{n-1}(\U_a \setminus \{u\} )}  \prod\limits_{v\in \tilde{\U}_a}w_{a,v} & n>1
\end{cases}
\end{align}
then the expression for probability $p_{a,u,n}$ simplifies to
\begin{align}
p_{a,u,n}=q_{a,u,n}p_{a,u,1}, \ n=1,2,\dots
\end{align} 
Note that $\sum_{n=1}^{|\U_a|}p_{a,u,n}=p_{a,u,1}\sum_{n=1}^{|\U_a|}q_{a,u,n} = 1$ and so $\sum_{n=1}^{|\U_a|}q_{a,u,n} =\frac{1}{p_{a,u,1}}$.

\subsection{802.11 MAC slots}
The number of users actively using an AP varies from time slot to time slot.   We therefore have to take some care to define the MAC slots within each WLAN appropriately.   Within each WLAN MAC slots are induced by carrier-sensing, which may be idle slots of duration $\sigma$ (no user transmits) or busy slots of duration $T_b$ (at least one user transmits, this includes both successful transmission and collision times).   In order to distinguish between MAC slots associated with different access points, we index MAC slots by $i \in \A \times \mathop{\mathbb{N}}$ i.e. the MAC slots associated with access point $a$ are $i=(a,1),(a,2),\dots$.   

\begin{figure}[h]
\centering
\begin{tikzpicture}
\node (t1) at (0,0) [draw,thick,minimum width=1.5cm,minimum height=.3cm] {};
\node (t2) at (1.5,0) [draw,thick,minimum width=1.5cm,minimum height=.3cm] {};
\node (tn) at (3.75,0) [draw,thick,minimum width=3cm,minimum height=.3cm] {};
\node (t) at (6,0) [draw,thick,minimum width=1.5cm,minimum height=.3cm] {};
\foreach \pos/\text in {{0.1,0.01}/$1$,
{1.6,0.01}/$2$,
{3.1,0.01}/$\dots$,
{6.1,0.01}/$t$} {
\draw (\pos) node {\text};
}
\draw[<->] (-0.75,+0.3) to node[pos=0.3,above] {$T$} (0.75,+0.3);
\draw[<->] (+0.75,+0.3) to node[pos=0.3,above] {$T$} (2.25,+0.3);
\draw[<->] (5.2,+0.3) to node[pos=0.3,above] {$T$} (6.7,+0.3);
\draw (-0.75,-0.2) -- (-1.2,-.6);
\draw (.75,-0.2) -- (1.2,-.6);
\node[rectangle split, rectangle split horizontal, rectangle split parts=10,draw,inner sep=.3ex] (MAC) at (0,-0.8)
{};
\node at (0,-1.1) [xscale=7,yscale=1.5,rotate=90] {$\{$};
\foreach \pos/\text in {{0.2,-1.4}/$\M_{a,1}$} {
\draw (\pos) node {\text};
}
\end{tikzpicture}
\caption{Illustrating 802.11 time slotting : each user $u$ can hop between multiple APs, but is associated to the same AP for time slots of duration $T$.  Within a time slot of duration $T$, 802.11 carrier sense defines MAC slots.  $\M_{a,t}$ denotes the set of MAC slots fully contained within time slot $t$.}
\label{fig:slots}
\end{figure}
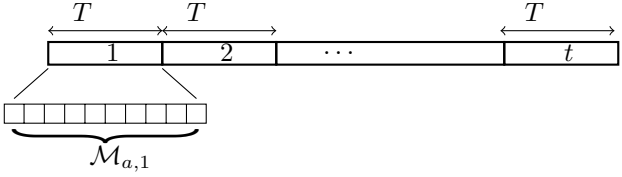

We let $\M_{a,t}$ denote the set of MAC slots of access point $a$ that are fully contained within time slot $t$ and $\bar{\M}_{a,t}$ be the set of MAC slots which are only partially contained in time slot $t$. Since the MAC slot duration is random, generally there will be boundary MAC slots which are only partially contained in time slot $t$ but there are at most two such slots.  We will assume that the time slot duration $T$ is sufficiently large that partial MAC slots $\bar{\M}_{a,t}$ can be neglected when calculating user airtime and throughput.  This is illustrated in Figure \ref{fig:slots}.

%


\subsection{802.11 throughput}
Let $L_{a,u}$, $T_{succ,a,u}$, $T_{coll}$ and $\sigma$ denote, respectively, the payload in bits of a successful transmission by station $u$ when associated with AP $a$, the mean duration of a successful transmission, the mean duration of a collision and the idle slot duration. 
To simplify notation we confine consideration to client uplink transmissions.  Extension to include AP downlink transmissions is straightforward, simply requiring appropriate book-keeping of which fraction of AP transmissions is directed to each client.   For simplicity we also assume that $T_{succ,a,u}$ is the same for all users transmitting in WLAN $a$ and let $T_{succ,a} = T_{succ,a, u}$.

In time slot $t$ let ${\U}_{a,t}\subset \U_a$ denote the set of users associated with access point $a$ and recall that $\mathbf{N}_{a,t}=|{\U}_{a,t}|$.  Let ${\X}_{i,u}$ be a random variable which is equal to $1$ if  station $u$  transmits in MAC slot $i\in\M_{a,t}$ and $0$ otherwise where $u \in {\U_a }$.   We let $\mathbf{\X}_{i,u}=0$ for users $u \notin {\U}_{a,t} $ not associated with access point $a$ in slot $i\in\M_{a,t}$.   For $u\in {\U}_{a,t}$ we assume that the $\{\mathbf{\X}_{i,u}\}$ are i.i.d. in both $u$ and $i$ i.e. $\mathbf{\X}_{i,u}\sim \mathbf{\X}$, and we let $\tau:=Prob(\mathbf{\X}=1)$.   This can be achieved in practice by setting $CW_{max}=CW_{min}$ and using the same value of $CW_{min}$ for all associated users, in which case $\tau=2/(CW_{min}+1)$.   
We let $\mathbf{\Y}_{i,u}$ be a random variable that equals 1 if station $u$ transmits successfully in MAC slot $i$, and $0$ otherwise. We let $\mathbf{\Z}_{i}$ equal $0$ when one or more users transmits in MAC slot $i$, and $1$ otherwise.  Since transmissions $\{\mathbf{\X}_{i,u}\}$ are i.i.d. for $u\in\U_{a,t}$, $i\in\M_{a,t}$, we have that $\mathbf{\Y}_{i,u}\sim \mathbf{\Y}^{|{\U}_{a,t}|}$ with $E[\mathbf{\Y}^{|{\U}_{a,t}|}]=\tau(1-\tau)^{|{\U}_{a,t}|-1}$, and $\mathbf{\Z}_{i}\sim \mathbf{\Z}^{|{\U}_{a,t}|}$ with $E[\mathbf{\Z}^{|{\U}_{a,t}|}]=(1-\tau)^{|{\U}_{a,t}|}$.  

The throughput of user $u$ in WLAN $a$ is  given by
\begin{align}\label{eq:air}
&s_{a,u} =\lim_{k\rightarrow\infty}\frac{\sum_{t=1}^k\sum_{i\in \M_{a,t}} \mathbf{\Y}_{i,u} L_{{a,u}}}{kT}\\
&= \lim_{k\rightarrow\infty}\frac{1}{k}\sum_{t\in\{s\in\{1,2,\dots,k\}:\mathbf{A}_{u,s}=a\}}\sum_{i\in \M_{a,t}} \mathbf{\Y}_{i,u}\frac{ L_{{a,u}}}{T} \\
&=  \lim_{k\rightarrow\infty}\sum_{n=1}^{|\U_a|}\frac{|\T_{a,n}^k|}{k}\frac{1}{|\T^k_{a,n}|}\sum_{t\in\T_{a,n}^k}\sum_{i\in \M_{a,t}} \mathbf{\Y}_{i,u}\frac{ L_{{a,u}}}{T}
\end{align}
where $\T_{a,n}^k:=\{s\in\{1,2,\dots,k\}:A_{u,s}=a,\mathbf{N}_{a,s}=n\}$ and we have used the fact that $\mathbf{x}_{i,u}=0$ when $u \notin { \U_a }$.  

Defining
\begin{align}
P_{succ,n}&=\sum_{j=1}^{n} E[\mathbf{\Y}^{n}] = n \tau (1-\tau)^{n-1}  \label{eq:P_Succn} \\
P_{idle,n}&=E[{\mathbf{\Z}^{n}}]=(1-\tau)^{n} \label{eq:P_idlen}\\
P_{coll,n}&=1-P_{idle,n}-P_{succ,n}
\end{align} 
we have
\begin{align}
&\lim_{k\rightarrow\infty}\frac{1}{|\T_{a,n}^k|}\sum_{t\in\T_{a,n}^k}\sum_{i\in \M_{a,t}} \mathbf{\Y}_{i,u} \frac{L_{{a,u}}}{T}  \\ 
&= E[|\M_{a,t}| |\mathbf{N}_{a,t}=n,\mathbf{A}_{u,t}=a] E[\mathbf{\Y^{n}}]\frac{L_{{a,u}}}{T} \\
&=\frac{\tau (1-\tau)^{n-1} L_{{a,u}} }{P_{idle,n}\sigma+P_{succ,n} T_{succ,a}+ P_{coll,n}T_{coll}} 
\end{align}
We also have $\lim_{k\rightarrow\infty}\frac{|\T_{a,n}^k|}{k} = Prob(\mathbf{A}_{u,t}=a,\mathbf{N}_{a,t}=n) = Prob(\mathbf{A}_{u,t}=a)Prob(\mathbf{N}_{a,t}=n|\mathbf{A}_{u,t}=a) = z_{a,u}p_{a,u,n}$.  
 Thus, the throughput of user $u$ in the WLAN operated by AP $a$ is given by
 \small
\begin{align}
s_{a,u} = z_{a,u}\sum_{n=1}^{|\U_a|}
&  \frac{  p_{a,u,n} \tau (1-\tau)^{n-1}  {L_{a,u}}}{P_{idle,n}\sigma+P_{succ,n} T_{succ,a}+ P_{coll,n}T_{coll}} \label{eq:80211T0}
\end{align}
\normalsize

\subsubsection{Each User Associates to a Single WLAN}
It is helpful to consider the special case where each user $u$ connects only to a single WLAN $a$, in which case $z_{a,u}=1$ and $z_{a^\prime,u}=0$ for $a^\prime\ne a$.   Letting $n_a$ denote the number of users connected to WLAN $a$ then $p_{a,u,n_a}=1$ and $p_{a,u,n}=0$ for $n\ne n_a$.   It follows that (\ref{eq:80211T0}) simplifies to,
\begin{align}
s_{a,u} =  
&   \frac{  \tau (1-\tau)^{n_a-1}  {L_{a,u}}}{P_{idle,n_a}\sigma+P_{succ,n_a} T_{succ,a}+ P_{coll,n_a}T_{coll}} 
\end{align}
which is identical to classical 802.11 WLAN models \emph{e.g.} that in \cite{leith2011log}.   However, when the number of users using a WLAN varies from time slot to time slot, (\ref{eq:80211T0}) yields the shared throughput.

\subsubsection{Change of variables}
It will prove useful to work in terms of the quantities $\phi=\frac{\tau}{1-\tau}$ and $w_{a,u}=\frac{z_{a,u}}{1-z_{a,u}}$ rather than $\tau$ and $z_{a,u}$s respectively.  Also we will use the normalised throughput $\rho_{a,u}=\frac{s_{a,u}}{c_{a,u}}$, where $c_{a,u}=\frac{L_{a,u}}{T_{s_{a}}}$.   Recalling $p_{a,u,n}=q_{a,u,n} p_{a,u,1}$ and  letting $N_{a}=\frac{T_{s_{a}}}{T_c}$, we have
\small
\begin{align}
\rho_{a,u}=   \frac{w_{a,u}}{1+w_{a,u}}\sum_{n=1}^{|\U_a|}\frac{q_{a,u,n}}{\prod\limits_{v\in \U_a\setminus\{u\}} (1+w_{a,v})} \frac{\phi}{\Phi_n(\phi)}  \frac{T_{s_{a}}}{T_c} \label{eq:80211T}
\end{align}
\normalsize
with 
\begin{align}
\Phi_n(\phi)=\frac{\sigma}{T_c}+n (N_{a}-1) \phi+ (\phi+1)^{n}-1
\end{align}
\subsubsection{802.11 Rate Region}
With the above change of variables the rate region of the 802.11 network is, 
\begin{align}
&\R_{wifi}=\bigg\{\mathbf{s}:   s_u=\sum_{a \in \A_u} \rho_{a,u} c_{a,u}, \nonumber\\
&\quad \rho_{a,u} \leq  \sum_{a \in \A_u}  \frac{w_{a,u}}{\prod\limits_{v\in \U_a} (1+w_{a,v})}\sum_{n=1}^{|\U_a|}q_{a,u,n}\frac{ \phi}{\Phi_n(\phi)} \frac{T_{s_ {a}}}{T_c}, \nonumber \\
&\quad q_{a,u,n} \leq \sum\limits_{\stackrel{\tilde{\U}_a\in}{\mathcal{P}_{n-1}(\U_a \setminus \{u\} )}}  \prod\limits_{v\in \tilde{\U}_a}w_{a,v},~ n > 1 ,\nonumber \\
 & \quad \sum_{a \in \A} \frac{w_{a,u}}{1+w_{a,u}} \leq 1,  0 \leq \rho_{a,u} \leq 1, \ w_{a,u} \ge 0,~ q_{a,u,n} \ge 0 \nonumber \\
& \hspace{3cm}  \forall n \in \U,~ \forall a \in \A,  ~\forall u \in \U \bigg\}
\end{align}
where $\mathbf{s}\in\mathbb{R}^{U}$ is the vector obtained by stacking the user throughputs $s_{u}$, $u\in\U$.
Letting $\tilde{\rho}_{a,u} =\log {\rho}_{a,u}$, {$\tilde{w}_{a,u}=\log {w_{a,u}}$},  we also have that the rate region of the 802.11 network can be written as
\begin{align}
&\hat{\R}_{wifi}=\bigg\{\mathbf{s}: {s}_{u}= \sum_{a \in \A_u} e^{\tilde{\rho}_{a,u}} c_{a,u} 
, \nonumber \\
&\tilde{\rho}_{a,u} \leq \tilde{w}_{a,u}-\sum_{v\in \U_a} \log(1+e^{\tilde{w}_{a,v}}) + \log \sum_{n=1}^{|\U_a|} \frac{T_{s_{a}}}{T_c}\frac{\phi}{\Phi_n}q_{a,u,n}, \nonumber \\
& q_{a,u,n} \leq \sum\limits_{\stackrel{\tilde{\U}_a\in}{\mathcal{P}_{n-1}(\U_a \setminus \{u\} )}}  \prod\limits_{v\in \tilde{\U}_a}e^{\tilde{w}_{a,v}},~ n > 1 \nonumber \\
& \sum_{a \in \A} \frac{e^{\tilde{w}_{a,u}}}{1+e^{\tilde{w}_{a,u}}} \leq 1,\ q_{a,u,n} \ge 0, \nonumber \\
& \hspace{3cm} \forall n \in \U,~ \forall a \in \A,~  \forall u \in \U \bigg\}
\label{eq:Rtildewifi}
\end{align}
\subsubsection{Example 802.11 Rate Region }
Recall the simple example in Fig \ref{fig:intro_example} where we have one BS, one AP and two users.   The  802.11 rate region is,

\begin{small}
\begin{align}
&\R_{wifi}=\bigg\{\mathbf{s}: s_{1} =  {{\rho}_{1,1}} c_{1,1},~ s_{2} =  {{\rho}_{1,2}} c_{1,2} \nonumber \\
& \quad {\rho}_{1,1} \leq \frac{w_{1,1}}{ (1+w_{1,1}) (1+w_{1,2})}(\frac{\phi}{\Phi_1}+q_{1,1,2}\frac{\phi}{\Phi_2}) \frac{T_{s_ {1,1}}}{T_c}, \nonumber \\
& \qquad q_{1,1,2} \leq w_{1,2},  \nonumber \\
&\quad {\rho}_{1,2} \leq \frac{w_{1,2}}{ (1+w_{1,1}) (1+w_{1,2})}(\frac{\phi}{\Phi_1}+q_{1,1,2}\frac{\phi}{\Phi_2}) \frac{T_{s_ {1,2}}}{T_c}, \nonumber \\
& \qquad  q_{1,2,2} \leq w_{1,1}, \nonumber \\
& \quad 0 \leq \rho_{1,u} \leq 1, \ w_{1,u} \ge 0,~ q_{1,u,n} \ge 0, ~ \big\{u,n| u,n \in \{1,2\}\big\}
\bigg\}
\label{eq:2usersRR}
\end{align}
\end{small}
\begin{figure}
\centering
    \subfloat[t][1 AP: $\tau_{a_1}=0.95$]{
    {\includegraphics[width=0.45\columnwidth]{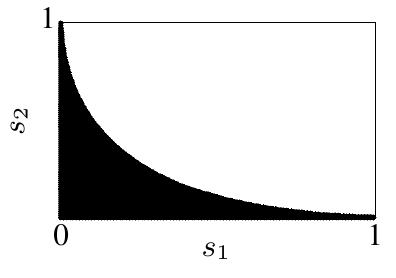}} \label{fig:rr2APsboth}
    }      
    \subfloat[t][2 APs: $\tau_{a_1}=0.95, \tau_{a_2}=0.2$]{
    {\includegraphics[scale=1]{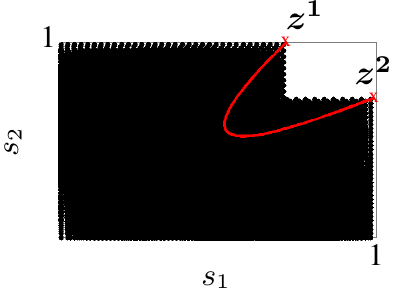}} \label{fig:rr2APs952}
    }     \\ 
\caption{Examples of 802.11 rate region $\R_{wifi}$ where hopping between APs is admissible.  In (a) two users hop between connecting and not connecting to an AP and in (b) two users hop between two APS for which the transmission probabilities differ.}\label{fig:hoppingRegion}
\end{figure}
%
%
%
\subsection{Non-convexity of Network Rate Region}

The network rate region is the joint LTE/802.11 rate region $\R_{lte}\times\R_{wifi}$.   The LTE rate region $\R_{lte}$ is convex and so the convexity/non-convexity of the network rate region depends on the convexity/non-convexity of the 802.11 component of the network rate region $\R_{wifi}$.   

In the special case (see above)  where for each user $u$ we force $z_{a,u}=1$ for one AP $a$, then each user associates to a single, fixed AP and the only design freedom lies in the LTE allocation which is a convex problem.  However, when we additionally allow the UEs to adjust the fraction of time that they connect to this single AP then the rate region becomes non-convex, as illustrated in Figure \ref{fig:hoppingRegion}(a), and when users may connect to more than one AP is illustrated in Fig \ref{fig:hoppingRegion}(b).  Fig \ref{fig:hoppingRegion}(b) plots the rate region $\R_{wifi}$ for networks with two APs and two UEs.   The non-convexity of the rate region in Fig \ref{fig:hoppingRegion}(b) is evident.   Note that the two extreme points of the ``notch'' in the rate region indicated in the figure correspond to parameter values $\mathbf{z}^1=[1~ 0~  0~ 1]^T$, and $\mathbf{z}^2=[0 ~1 ~1 ~0]^T$, where $\mathbf{z}:=[z_{1,1}~ z_{1,2}~ z_{2,1}~ z_{2,2}]^T$.  That is, point $\mathbf{z}^1$ corresponds to $u_1$ being connected to AP $a_1$ 100\% of the time and  $u_2$ to AP $a_2$, while $\mathbf{z}^2$ is the reverse.  The UE rates along the time-sharing chord $\alpha\mathbf{z}^1+(1-\alpha)\mathbf{z}^2$, $\alpha\in[0,1]$ are indicated by the line marked on Fig \ref{fig:hoppingRegion}(b).   It can be seen that this lies in the \emph{interior} of the rate region.  Simple time-sharing therefore does not yield convexity, and this is due to the collisions in 802.11 WLANs shared by more than one UE.


\section{Proportional Fair Rate Allocation}\label{sec:optimisation}
\subsection{Utility Fair Optimisation}
We are now in a position to consider the main focus of this work, namely finding a proportional fair rate allocation for the joint LTE/802.11 network.   This is the solution to the following utility optimisation $P$, 
\begin{align}
&\max_{\mathbf{s},\mathbf{r}} 
\sum_{u\in \U} \log \left(s_u + r_u\right)\\
s.t. \quad 
&\mathbf{s} \in \hat{\R}_{wifi},\ \mathbf{r}\in\R_{lte}
\end{align}

As already noted, the rate region ${\R}_{wifi}$ (and so $\hat{\R}_{wifi}$) is generally non-convex and so the utility fair optimisation $P$ is non-convex.   However, as we will see the optimisation possesses sufficient convex structure to allow near-optimal solutions to be found in an efficient manner.  We proceed by first considering methods for approximating a non-convex set by a maximal convex subset.  In this way we can define a convex approximation to $P$ for which solutions can be efficiently found.   We then consider a concave-convex procedure that adaptively selects the maximal convex subsets with the aim of maximising the optimisation objective.   This procedure is guaranteed to converge to a stationary point of non-convex optimisation $P$.  Although this stationary will in general be sub-optimal, in practice we find that it is usually near-optimal.

\subsection{Approximate Optimisation Via Maximal Convex Subsets}
\subsubsection{Maximal convex subsets}\label{subsec:MCS}

\begin{figure}
\centering
\includegraphics[width=0.6\columnwidth]{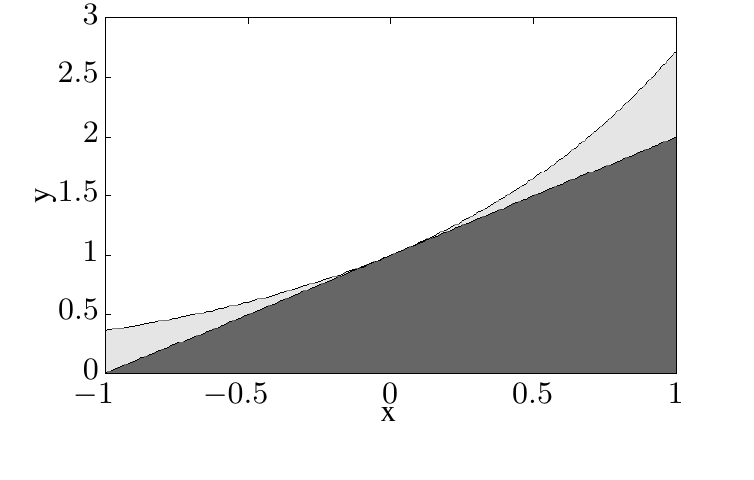}
\caption{Illustrating a maximal convex subset (indicated in darker shade) of concave set $y\le e^x$.}\label{fig:concave}
\end{figure}

Sets of the form $E:=\{(x,y): y\le e^x, (x,y)\in\mathbb{R}^2\}$ will prove important in our analysis. Set $E$ is concave, see Figure \ref{fig:concave}.  Nevertheless, large convex subsets of this set can be readily identified.   In particular, the set $F:=\{(x,y):y\le e^{\bar{x}}(1+x-\bar{x}), (y,x)\in\mathbb{R}^2\}$, with parameter $\bar{x}\in\mathbb{R
}$, is convex with $F\subset E$, as illustrated in Figure \ref{fig:concave}.   Convexity follows from the fact that the constraint defining set $F$ is linear in $x$ and $y$.   That $F$ is contained in $E$ follows from the observation that the complement $\bar{E}:=\{(x,y): y\ge e^x, (x,y)\in\mathbb{R}^2\}$ of set $E$ is convex and the boundary of $F$ is a supporting hyperplane to $\bar{E}$ at point $(x,e^{\bar{x}})$.

\subsubsection{Maximal Convex Subsets of 802.11 Rate Region}
The main constraints in the 802.11 rate region $\hat{\R}_{wifi}$ can be written in standard form as
\begin{align}
&\tilde{\rho}_{a,u} 
-  \tilde{w}_{a,u}+\sum_{v\in \U_a} \log(1+e^{\tilde{w}_{a,v}})   - \log \left(\sum_{n=1}^{|\U_a|} \frac{\phi}{\Phi_n}q_{a,u,n} \right)  \le 0 \label{eq:c2} \\
&q_{a,u,n} - \sum\limits_{\stackrel{\tilde{\U}_a\in}{\mathcal{P}_{n-1}(\U_a \setminus \{u\} )}}  \prod_{v\in \tilde{\U}_a}e^{\tilde{w}_{a,v}}\le 0 \label{eq:c3}
\end{align}
 The terms on the LHS of the $\tilde{\rho}_{a,u}$ constraint are all convex.  However, the second term on the LHS of the $q_{a,u,n}$ constraint is concave.  We proceed by finding a maximal convex subset.

Letting
\begin{align}
E:=\{(&q_{a,u,n}, \{\tilde{w}_{a,v}\}_{v\in\U_a}): \tilde{w}_{a,v}\in\mathbb{R}, v\in\U_a,\notag\\
&q_{a,u,n}\le \sum\limits_{\stackrel{\tilde{\U}_a\in}{\mathcal{P}_{n-1}(\U_a \setminus \{u\} )}}  \prod_{v\in \tilde{\U}_a}e^{\tilde{w}_{a,v}}\}
\end{align}
then a maximal convex subset is
\begin{align}
F:= \{(&q_{a,u,n}, \{\tilde{w}_{a,v}\}_{v\in\U_a}):   \tilde{w}_{a,v}\in\mathbb{R}, v\in\U_a, \notag\\
&q_{a,u,n}\le \sum\limits_{\stackrel{\tilde{\U}_a\in}{\mathcal{P}_{n-1}(\U_a \setminus \{u\} )}}\prod_{v\in \tilde{\U}_a}W_{\tilde{\U}_a,v}
\label{eq:chat3}
\end{align}
where $W_{\tilde{\U}_a,v}:=  e^{\bar{w}_{a,v}} (1+ \sum_{v\in \tilde{U}_a\setminus\{u\}}(\tilde{w}_{a,v}-\bar{w}_{a,v}))$ and $\bar{w}_{a,u}$ is a design parameter.

{The 802.11 rate region also contains the constraint 
\begin{align}
 s_u - \sum_{a \in \A_u} e^{\tilde{\rho}_{a,u}} c_{a,u} \leq  0
 \end{align}
It can be seen that the LHS is concave. Once again, we adopt a maximal convex subset approach $$ s_u  - \sum_{a \in \A_u} e^{\bar{\rho}_{a,u}}(1+\tilde{\rho}_{a,u}-\bar{\rho}_{a,u})c_{a,u} \leq 0$$ }
where $\bar{\rho}_{a,u}$ is a design parameter.  The LHS is now convex in $\tilde{\rho}_{a,u}$  as required.


Lastly, the 802.11 rate region contains the constraint
\begin{align}
\sum_{a \in \A} \frac{e^{\tilde{w}_{a,u}}}{1+e^{\tilde{w}_{a,u}}} \leq 1
\label{eq:c4}
\end{align}
the LHS of which is sigmoidal in form, so neither convex not concave.
%
Rewriting $\frac{e^{\tilde{w}_{a,u}}}{e^{\tilde{w}_{a,u}}+1}$ equivalently as $e^{\tilde{w}_{a,u}} - \frac{e^{2\tilde{w}_{a,u}}}{1+e^{\tilde{w}_{a,u}}}$ observe that the first term in the RHS is convex while the second term is concave. Linearising the second term yields the following inequality
\begin{align}
\sum_{a \in \A}e^{\tilde{w}_{a,u}} - \left(\frac{2 e^{2\bar{w}_{a,u}}}{1+e^{\bar{w}_{a,u}}}+\frac{e^{3\bar{w}_{a,u}}}{({1+e^{\bar{w}_{a,u}}})^2}\right) (\tilde{w}_{a,u}-\bar{w}_{a,u})\le 1
\label{eq:chat4}
\end{align}
which defines a maximal convex subset.




\subsubsection{Solving the Convex Optimisation}
Using the maximal convex subset approach described above we obtain a convex optimisation the solution of which is feasible for non-convex optimisation $P$ but will, in general, be sub-optimal.   The Slater condition is satisfied and so strong duality holds.   We can therefore solve this convex optimisation in a distributed manner using a primal-dual subgradient approach. 

{Let $\mathbf{x}:=[\boldsymbol{\zeta}^T~ \mathbf{r}^T ~\mathbf{s}^T~ \boldsymbol{\tilde{\rho}}^T ~\mathbf{\tilde{w}}^T~ \mathbf{q}^T]^T\in\mathbb{R}^n$  be the vector obtained by stacking LTE and 802.11 rate region variables with $\boldsymbol{\zeta}\in \mathbb{R}^{|\I|\times |\B|\times U}$, $\boldsymbol{\tilde{\rho}},\mathbf{\tilde{w}}\in \mathbb{R}^{|\A|\times U}$, $\mathbf{q}\in \mathbb{R}^{|\A|\times U\times U}$ denoting, respectively, the vectors with elements  $\zeta^i_{b,u}$, $\tilde{\rho}_{a,u}$, $\tilde{w}_{a,u}$, $q_{a,u,n}$ for $b \in \B$, $i \in \I$, $a \in \A$, $n \in \U$, $ u \in \U$. 
We re-write the optimisation problem $P$ in the  following form,
\begin{align}
&\min_{\mathbf{x}} ~ {f}(\mathbf{x})\\
&s.t. \quad {h}^{(i)}(\mathbf{x})-{g}^{(i)}(\mathbf{x}) \leq \mathbf{0}, ~ i=1,2,\dots,l
\end{align}
with $f:\mathbb{R}^n\rightarrow\mathbb{R}$, $h^{(i)}:\mathbb{R}^n\rightarrow\mathbb{R}$, $g^{(i)}:\mathbb{R}\rightarrow\mathbb{R}$ being convex continuously differentiable functions, expressions for which are given in  Appendix A. {Let $-{\hat{g}}^{(i)}(\mathbf{x};{\bar{\mathbf{x}}} )$ be the maximal convex subset expression for non-convex function $-g^{(i)}(\mathbf{x} )$.
\begin{align}
-{\hat{g}}^{(i)}(\mathbf{x};{\bar{\mathbf{x}}} )=-g^{(i)}(\bar{\mathbf{x}})-\partial g^{(i)}_{\mathbf{x}}(\bar{\mathbf{x}}) (\mathbf{x}-\bar{\mathbf{x}})
\end{align}
Then the approximate  optimisation problem $P_{\bar{\mathbf{x}}}$ is given by
\begin{align}
&\min_{\mathbf{x}} ~ {f}(\mathbf{x})\\
&s.t. \quad {h}^{(i)}(\mathbf{x})-{\hat{g}}^{(i)}(\mathbf{x};{\bar{\mathbf{x}}} ) \leq \mathbf{0}, ~ i=1,2,\dots,l
\end{align}
}
Letting $\Lambda:=[\lambda^{(1)},\cdots, \lambda^{(l)}]^T$ denote the set of multipliers associated with the rate region constraints 1 to $l$, the Lagrangian for optimisation problem $P_{\bar{\mathbf{x}}}$ is}
%
%
\begin{eqnarray*}
L(\mathbf{x},\Lambda; \mathbf{\bar{\mathbf{x}}})&=&f( \mathbf{x})
 +\sum_{i=1}^l \lambda^{(i)} \left(h^{(i)}( \mathbf{x})-\hat{g}^{(i)} (\mathbf{x}; \mathbf{\bar{\mathbf{x}}}) \right)
\end{eqnarray*}
The standard primal-dual subgradient approach in Algorithm \ref{algo1} can then be used, for example, to find a solution to optimisation $P_{\bar{\mathbf{x}}}$.
%
\begin{algorithm}[!htb]
\footnotesize
Initialise: $t=0$, $\mathbf{x}(0)$, $\Lambda(0)$, step size $\alpha>0$\\
\textbf{do}\\
\begin{minipage}[c]{\columnwidth}
\begin{align}
 %
&\mathbf{x}(t+1) = \mathbf{x}(t) - \alpha \partial_{\mathbf{x}}L\big(\mathbf{x}\left(t\right), \Lambda\left(t\right);\bar{\mathbf{\mathbf{x}}} \big) \nonumber \\
&\Lambda(t+1) = \left[\Lambda(t) + \alpha \partial_{\Lambda}L\big(\mathbf{x}\left(t\right),\Lambda\left(t\right); \bar{\mathbf{\mathbf{x}}}\big)\right]^+  \nonumber \\
&t\leftarrow t+1 \nonumber
\end{align}
\end{minipage}
\textbf{loop}
\caption{Distributed primal-dual algortithm}\label{algo1}
\end{algorithm}
%

%
%

\subsubsection{Message Passing Required}
\begin{enumerate}
\item To update the LTE sub-channel airtime $\zeta^i_{b,u}$ each UE needs (i) the SNR to its LTE BSs (which it already knows) and (ii) the multiplier associated with the sub-band constraint (which can be communicated by a BS).
\item The 802.11 association probability $\tilde{w}_{a,u}$ and WLAN parameters $q_{a,u,n}$ can be updated using information available locally at AP $a$ (no need for message passing) together with knowledge of the multiplier associated with the constraint that association probabilities for user $u$ sum to one.  This requires that $\tilde{w}_{a,u}$  {as well as $\bar{w}_{a,u}$} be communicated by AP $a$ to UE $u$ and the multiplier then communicated back from UE $u$ to  AP $a$.    
\item The 802.11 normalised throughput $\tilde{\rho}_{a,u}$ can be updated by UE $u$ using local information together with knowledge of the multiplier for the rate constraint on $\tilde{\rho}_{a,u}$ (which can be communicated by AP $a$).
\end{enumerate}

\subsection{Adaptation of Maximal Convex Subsets }
The convex optimisation in the preceding section yields a solution which is feasible for non-convex optimisation $P$ but which is, in general, sub-optimal for $P$.    The degree of sub-optimality is dependent on the choice of maximal convex subsets used to derive the convex optimisation, \emph{e.g.} when the convex subsets contain at least one point which is an optimum of the non-convex problem $P$ then the solution to the convex optimisation $P_{\bar{\mathbf{x}}}$ will in fact be optimal for $P$.   

When we have some knowledge of the part of the network rate region where the optima of problem $P$ are likely to lie, we can use this to select the maximal convex subsets.   Such knowledge might be available from prior experience, \emph{e.g.} previous solutions to similar network configurations, or from structural insight.   In the next section we illustrate how such prior information can be used to obtain near-optimal solutions.

However, it is also possible to automate this process and this is the focus of the present section.  The basic idea is to  iteratively update the choice of maximal convex subsets based on the current solution to the convex optimisation.  Having obtained the solution to the convex optimisation for the current choice of maximal convex subsets, one natural approach is to use the components $\tilde{w}_{a,u}$, $\tilde{\rho}_{a,u}$ of this solution as the values for the parameters $\bar{w}_{a,u}$, $\bar{\rho}_{a,u}$ and in this way to define a new maximal convex subset.   
%
%

\subsubsection{Convergence}
In more detail, let $C:=\{\mathbf{x}\in B:h^{(i)}(\textbf{x})-g^{(i)}(\textbf{x})\le 0,\ i=1,\cdots.,l\}$ with $B\subset \mathbb{R}^n$ non-empty, convex and compact (closed and bounded).  As we will see, set $B$ is needed for technical reasons, to ensure that $C$ is compact, but $B$ can otherwise be chosen arbitrarily and can be viewed as augmenting the set of convex constraints $h^{(i)}$.  Of course, we assume that set $C$ is non-empty.  Now consider the iterative update
\begin{align}
&\mathbf{x}_{k+1} \in D_{\mathbf{x}_k} f(\mathbf{x}),\ k=1,\cdots \label{eq:a1}
\end{align}
with $\mathbf{x}_1\in C$,
\begin{align}
D_{\mathbf{x}_k}f(\mathbf{x})=
\begin{cases}
\arg\min_{\mathbf{x}\in C_{\mathbf{x}_k}} f(\mathbf{x}) & \text{if }  f(\mathbf{x}_k)=f^*_{\mathbf{x}_k}\\
\{\mathbf{x}\in C_{\mathbf{x}_k}:f(\mathbf{x}) < f(\mathbf{x}_k)\} & \text{otherwise}
\end{cases}
\end{align}
and $f^*_{\mathbf{x}_k}:=\min_{\mathbf{x}\in C_{\mathbf{x}_k}} f(\mathbf{x})$, $C_{\mathbf{x}_k}:=\{\mathbf{x}\in B:h^{(i)}(\mathbf{x}) - \hat{g}^{(i)}(\mathbf{x};\mathbf{x}_k) \le 0\}$.  

Observe that the solution to convex optimisation $P_{\mathbf{x}_k}$ lies in set $D_{\mathbf{x}_k} f(\mathbf{x})$ and so can be used in update (\ref{eq:a1}).  However, set $D_{\mathbf{x}_k} f(\mathbf{x})$ defines a wider class of updates, including suitable approximate solutions to $P_{\mathbf{x}_k}$ that may be easier/faster to find, and  generalises the concave-convex approach of \cite{yuille2003concave}.

We have the following convergence result.
\begin{theorem}[Local Convergence]\label{lem:one}
Iterative update (\ref{eq:a1}) converges to a stationary point of non-convex optimisation $P$.
\end{theorem}
\begin{proof}
See Appendix B.
\end{proof}


By adding a stochastic search component to update (\ref{eq:a1}) we can strengthen this to obtain a global convergence result.

\begin{theorem}[Global Optimisation] \label{lem:two}
Consider iterative update
\begin{align}
&\mathbf{x}_{k+1} \in 
\begin{cases}
\mathbf{X}_{k+1} & \text{w.p.}\ \epsilon\\
D_{\mathbf{x}_k} f(\mathbf{x}) & \text{otherwise} \label{eq:a2}
\end{cases} \nonumber \\
&y_{k+1} = \arg \min\{f(y_k), f(\mathbf{x}_{k+1})\}  
\end{align}
with $\mathbf{x}_1\in C$ and where $\{\mathbf{X}_{k}\}$ is a sequence of random variables  taking values in $C$ satisfying $Prob(\mathbf{X}_{k+1} \in \Sigma | \mathbf{X}_k \in \Upsilon) \ge \eta \mu(\Upsilon\cap \Sigma)$ for any $\Upsilon$, $\Sigma \subset C$ where $\mu(\Upsilon\cap \Sigma)$ denotes the volume of their intersection and parameter $\eta>0$.   Then, $y_{k}$ converges to the optimum of optimisation P with probability one. 
\end{theorem}
\begin{proof}
See Appendix B.
\end{proof}
Update (\ref{eq:a2}) is no longer purely a descent update, but rather with probability $\epsilon$ an update is made which may lead to the objective $f$ increasing, allowing escape from unfavourable stationary points (saddle points \emph{etc}) and from local minima.   The requirement on random variable $\mathbf{X}_{k+1}$ can be satisfied by, for example, selecting $\mathbf{X}_{k+1}$ uniformly at random within a ball about the current point $\mathbf{x}_k$.  

Although the proof of Theorem \ref{lem:two} could be used to upper bound the convergence, this bound would be very loose and so while Theorem \ref{lem:two} provides some comfort as to the ability of update (\ref{eq:a2}) to find a global optimum, evaluation of the convergence rate really needs to be carried out via numerical experiments.

{Note that keeping track of the best $x_{k+1}$ at each step of Update (\ref{eq:a2}) doesn't lend itself to distributed implementation and so we drop that step in the proceeding examples (although this also means that we lose the convergence result of theorem 2).}
\subsubsection{Descent Step}

Descent step $D_{\mathbf{x}_k}f(\mathbf{x})$ can be realised in many ways.  One is by use of a truncated version of the primal-dual subgradient approach:

\begin{algorithm}[h!]
\footnotesize
\begin{minipage}[c]{\columnwidth}
\begin{align*} 
&\textbf{do}\\
&\qquad \mathbf{z}(t) = \mathbf{x}(t)\\
&\qquad \mathbf{z}(t+1) = \mathbf{z}(t+1) - \alpha\partial_{\mathbf{z}}L\big(\mathbf{z}\left(t\right), \Lambda\left(t\right);{\mathbf{x}_k} \big)\\
&\qquad\Lambda^{(i)}(t+1) = \left[\Lambda^{(i)}(t) + \alpha \partial_{\Lambda}L\big(\mathbf{z}\left(t\right), \Lambda\left(t\right);{\mathbf{x}_k} \big)\right]^+\\
&\qquad t=t+1\\
&\textbf{until}\ f(\mathbf{z}(t+1)) \le f(\mathbf{x}_k) \text{ and } \mathbf{z}(t+1)\in C_{\mathbf{x}_k}\\
&\mathbf{x}_{k+1} =\mathbf{ z}(t+1)
\end{align*}
\end{minipage}
\caption{$D_{\mathbf{x}_k}f(\mathbf{x})$: Truncated Primal-Dual Updates}\label{algo:descentStep}
\end{algorithm}
%
%
%
While this primal-dual approach might not lead to a feasible point at every step, the requirement that $\mathbf{x}_{k+1}$ be feasible can be relaxed to the requirement that (i) $\mathbf{x}_{k+1}$ is bounded and (ii) $\mathbf{x}_{k+1}$ is feasible for sufficiently large $k$.  Theorem \ref{lem:one} then continues to hold, with only minor changes to the proof. 

\subsubsection{Example}
\begin{figure}
\subfloat[t][]{
\includegraphics[scale=1]{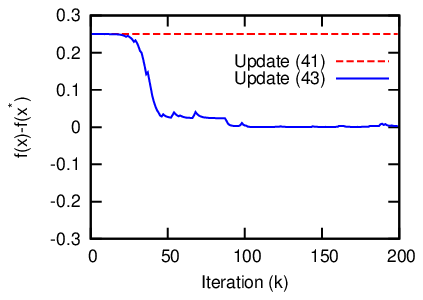} }
\subfloat[t][]  {
\includegraphics[scale=1]{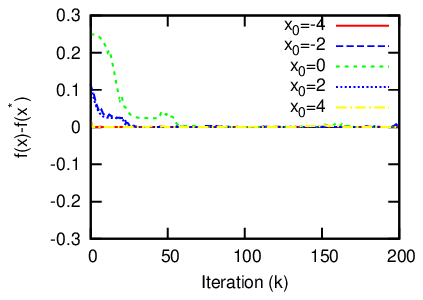} }
\caption{ Example illustrating global convergence, $\alpha=0.1$, $\epsilon=0.2$. }
\label{fig:globalConvg}
\end{figure}

The following simple example illustrates the convergence of update (\ref{eq:a2}) to a global optimum.  Consider the optimisation problem 
\begin{align*}
\min_{x\in[-4,4]} \quad x^2-x^4
\end{align*}
{It can be verified that in the interval $[-4,4]$ this has a global minimum  at $x^*= \pm 1/\sqrt{2}$ with $f(x^*)=-0.25$, but it also has a stationary point  at $x=0$.    Fig \ref{fig:globalConvg}(a) compares update (\ref{eq:a2}) with update (\ref{eq:a1}) when starting from initial condition $x=0$ (a stationary point).} It can be seen that update (\ref{eq:a1}) gets stuck at this stationary point whereas update (\ref{eq:a2}) is able to escape and find a global minimum.   Fig \ref{fig:globalConvg}(b)  shows realisations of update (\ref{eq:a2}) for a range of initial conditions, illustrating its insensitivity to the choice of initial condition.

%

\section{Motivating Example Revisited} \label{sec:simpleEx}
Armed with our proportional fair problem formulation and non-convex optimisation tools we now revisit the example in Fig \ref{fig:intro_example} in more detail.
  
\subsection{Network Setup}
\begin{figure}[!htb]
\centering
\begin{tikzpicture}
\node (PHYH) at (0cm,0cm) [draw,thick,minimum width=2cm,minimum height=.3cm] {};
\node (PSDU) at (3.75cm,0cm) [draw,thick,minimum width=5.5cm,minimum height=.3cm] {};
\foreach \pos/\text in {{0.06,0.01}/ \scriptsize PHY Header,
{4,0.01}/\scriptsize A-MPDU}{
\draw (\pos) node {\text};
}
\draw[<->] (+1,+0.3) to node[pos=0.3,above] {\scriptsize PSDU} (6.5,+0.3);
\draw (1,-0.2) -- (0.3,-0.6);
\draw (6.5,-0.2) -- (7.2,-0.6);
\node (sf1) at (1,-0.8) [draw,thick,minimum width=1.5cm,minimum height=.3cm] {};
\node (sf2) at (2.5,-0.8) [draw,thick,minimum width=1.5cm,minimum height=.3cm] {};
\node (sfn) at (4.5,-0.8) [draw,thick,minimum width=2.5cm,minimum height=.3cm] {};
\node (sfk) at (6.5,-0.8) [draw,thick,minimum width=1.5cm,minimum height=.3cm] {};
\foreach \pos/\text in {{1,-0.8}/\scriptsize Subframe 1,
{2.5,-0.8}/\scriptsize Subframe 2,
{4.5,-0.8}/$\dots$,
{6.5,-0.8}/\scriptsize Subframe K} {
\draw (\pos) node {\text};
}
\draw (1.75,-1) -- (.1,-1.4);
\draw (3.25,-1) -- (5.2,-1.4);
\node (deliimiter) at (.5,-1.7) [draw,thick,minimum width=1cm,minimum height=.5cm] {};
\node (t2) at (1.4,-1.7) [draw,thick,minimum width=0.8cm,minimum height=.5cm] {};
\node (tn) at (2.8,-1.7) [draw,thick,minimum width=2cm,minimum height=.5cm] {};
\node (t) at (4.2,-1.7) [draw,thick,minimum width=0.8cm,minimum height=.5cm] {};
\node (t) at (5,-1.7) [draw,thick,minimum width=0.8cm,minimum height=.5cm] {};
\foreach \pos/\text in {{.5,-1.7}/\scriptsize Delimiter,
{1.4,-1.6}/\scriptsize MPDU,
{1.4,-1.8}/\scriptsize Header,
{2.8,-1.7}/\scriptsize MSDU,
{4.2,-1.7}/\scriptsize FSC,
{5,-1.7}/\scriptsize Padding}  {
\draw (\pos) node {\text};
}
\end{tikzpicture}
\caption{802.11n frame format for Aggregated-Mac Packet Data Units (A-MPDUs).}
\label{fig:AMPDU_frame}
\end{figure}
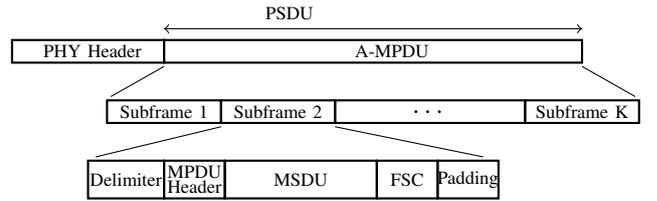

The network consists of one LTE base-station, one 802.11 AP and two UEs.    
We assume that the AP uses 802.11n \cite{IEEE802.11nstandard} with the settings detailed in Table \ref{tb:parameters} and the A-MPDU frame structure  illustrated in Fig \ref{fig:AMPDU_frame}.
%
%
For simplicity we assume that only two LTE sub-channels are available, denoted sub-band one and sub-band two. 
We consider LTE Frequency Division Duplex (FDD) systems with  10MHz system bandwidth and uplink transmissions with one data stream from single antenna UEs.  The LTE system parameters are also detailed in Table \ref{tb:parameters}. 

%

\begin{table}[!htb]
\caption{Network parameters.}
\label{tb:parameters}
\centering
\scalebox{0.65}
{
\begin{tabular}{|l|l|l|}
\hline
\multirow{4}{*}{802.11 Parameters}
&$CW_{MIN} $ & $15$\\ 
&Physical Rate, $(c_{a,1},c_{a,2})$& $(1,54)$ Mbps\\ 
&Physical Rate, $(c_{b,1},c_{b,2})$ & $(5,1)$ Mbps \\ 
& Basic Rate ($R_{basic}$) & $1~$Mbps \\ 
& Slot Time ($\sigma$) &$9 \mu s$\\ 
& PLCP Preamble& $16 \mu s$ \\ 
& PLCP Header & $48 ~$bits $@c_{a,u}$\\ 
& MAC Header & $ 192~$bits $@c_{a,u}$\\ 
& Payload size $(L)$ & $1500~$bytes $@c_{a,u}$\\ 
& Number of MPDU subframes for $u_1$ $(K_1)$ & $1~$\\ 
& Number of MPDU subframes for $u_2$ $(K_2)$ & $54~$\\
& FSC & $32~$bits $@c_{a,u}$ \\
& Delimiter & $32 ~$ bits $@c_{a,u}$\\
& Padding & $32 ~$ bits $@c_{a,u}$\\
& ACK & $ 112~$bits $@ R_{basic}$ \\ 
& RTS & $160 ~$bits $@ R_{basic}$  \\ 
& CTS & $112~$bits $@ R_{basic}$ \\ \hline
\multirow{4}{*}{LTE Parameters} 
& Duplex Mode & FDD \\
& System Bandwidth & $2\times10$ MHz \\
& Carrier Frequency & $2$ GHz \\
& Number of Antennas at BS & $1$ \\
& Number of Antennas at UE & $1$ \\
& Guard Band Overhead & $10\%$ \\
& DMRS Overhead & $14.3\%$ \\
& Random Access Overhead & $0.625\%$\\
& Cyclic Prefix Overhead & $6.66\%$ \\ 
& Bandwidth Efficiency ($\beta_1$) & $71\%$ \\
& SNR Efficiency ($\beta_2$) & $1$\\ 
& SNR ,$(\gamma^{(i)}_{1,1},\gamma^{(i)}_{1,2}),~ i=1,2$ & $(4.9, 4.9) ~dB$\\ \hline 
\multirow{2}{*}{Optimisation: ~
 \textbf{}}
&Maximal Convex Subset 1 ($\bar{\mathbf{x}}_1$) with $\{(\bar{\rho}_{a,1},\bar{\rho}_{a,2})$, $(\bar{w}_{a,1},\bar{w}_{a,2})\}$ & $\{(-1,0)$,$(-2,2)\}$  \\ & & \\
& Maximal Convex Subset 2 ($\bar{\mathbf{x}}_2$) with \{$(\bar{\rho}_{a,1},\bar{\rho}_{a,2})$,$(\bar{w}_{a,1},\bar{w}_{a,2})\}$ & $\{(0,0)$ ,$(0,0)\}$ \\ & & \\
&Maximal Convex Subset 3 ($\bar{\mathbf{x}}_3$) with $\{(\bar{\rho}_{a,1},\bar{\rho}_{a,2})$,$(\bar{w}_{a,1},\bar{w}_{a,2})\}$ & $\{(0,0)$,$(1,1)\} $\\& &  \\
&$\alpha$ (step size) & $0.01$\\ 
&$\epsilon$ & $0.2$\\
\hline
\end{tabular}
}
\end{table}

\subsection{Rate Allocations}
Using the maximal convex subset $\bar{\mathbf{x}}_1$ given in Table \ref{tb:parameters}, Algorithm \ref{algo1} yields rate allocations of 10 Mbps and  42.4Mbps for UEs $u_1$ and $u_2$ respectively.    Using update (\ref{eq:a1}) to adapt the maximal convex subset the rate allocation improves to be 10 Mbps for UE $u_1$ and 46.5Mbps for UE $u_2$.   Details of the solutions found using the two approaches are given in Table \ref{tb:simpleresults}.

For comparison, when $u_1$ and $u_2$ use only the 802.11 WLAN their data rates are, respectively 0.46Mbps and 24.9Mbps while when the UEs use only the LTE the BS allocates a rate of 5Mbps to each UE.   Splitting the traffic for each UE equally over the 802.11 and LTE networks would yield rates of 5.46Mbps and 29.9Mbps.

\begin{table}[!htb]
\caption{Solutions given by Algorithm \ref{algo1} and update (\ref{eq:a1}).}
\label{tb:simpleresults}
\centering
\begin{tabular}{|l|l|l|l|l|}
\hline
&\multicolumn{2}{c|}{Algorithm \ref{algo1}, Convex subset $\bar{\mathbf{x}}_1$} & \multicolumn{2}{c|}{Update (\ref{eq:a1})} \\ \cline{2-5}
\multicolumn{1}{|l|}{Parameter}&\multicolumn{1}{l|}{$u_1$}&\multicolumn{1}{l|}{$u_2$}&\multicolumn{1}{l|}{$u_1$}&\multicolumn{1}{l|}{$u_2$}\\
\hline\hline
$\zeta_{b,u}$ & $ 1$  &  $0$ & $1$ & $0$\\
$\rho_{a,u}$ & $0.05$   &    $0.85$ &   $0$ & $0.98$  \\
$r_{u}$ & $10$    &     $46.5$& $10$ &  $48.8$ \\
$p_{a,u,1}$ & $0.06$   &    $0.98$ & $0.01$& $0.99$ \\
$p_{a,u,2}$ & $ 0.85$   &     $0.01$& $0.97$& $0.03$ \\
$z_{a,u}$ & $0$  &      $0.93$& $0.01$ & $0.99$ \\
\hline
\end{tabular}
\end{table}
\subsection{Convergence}
\begin{figure}[!htb]
\centering
    \subfloat[t][Throughput of UE $u_1$]{
    {\includegraphics[scale=.95]{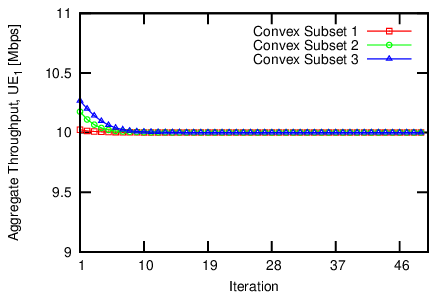}} \label{fig:Th1CCP}
    }      
\subfloat[t][Throughput of UE $u_2$]{
    {\includegraphics[scale=.95]{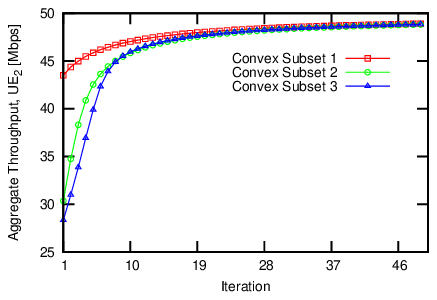}} \label{fig:Th2CCP}
   }\\
    \subfloat[t][${z_{a,1}}$]{
    {\includegraphics[scale=.95]{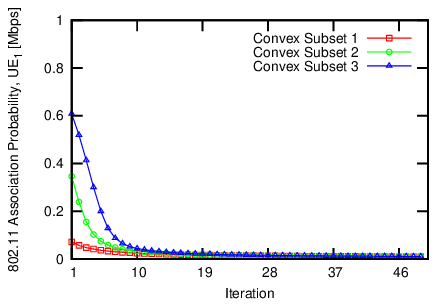}} \label{fig:z1CCP}
    }      
\subfloat[t][${z_{a,2}}$]{
    {\includegraphics[scale=.95]{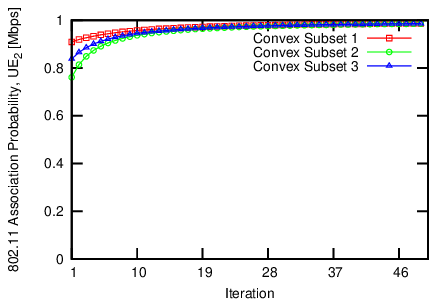}} \label{fig:z2CCP}
   }\\
\centering
    \subfloat[t][${\zeta^i_{b,1}},i=1,2$]{
    {\includegraphics[scale=.95]{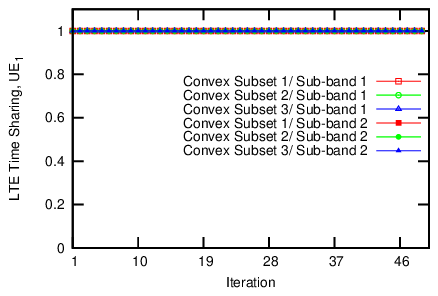}} \label{fig:lte1CCP}
    }      
\subfloat[t][${\zeta^i_{b,2}},i=1,2$]{
    {\includegraphics[scale=.95]{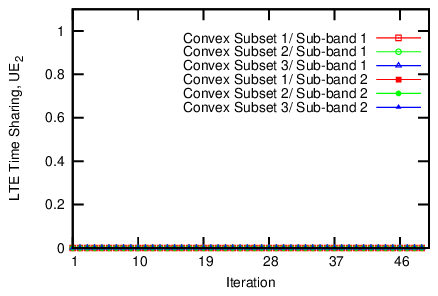}}\label{fig:lte2CCP}
   }\\
\caption{Convergence of update (\ref{eq:a1}) for various choices of initial convex subset, detailed in Table \ref{tb:parameters}. } \label{fig:Algo2Convergence}
\end{figure}

Figure \ref{fig:Algo2Convergence} illustrates the convergence of update (\ref{eq:a1}) for various choices of initial convex subset, detailed in Table \ref{tb:parameters}.  The first choice $\bar{\mathbf{x}}_1$ makes use of knowledge of the network to estimate the rate region in which the optimum is likely to lie.  The second and third choices $\bar{\mathbf{x}}_2$ and $\bar{\mathbf{x}}_3$ are randomly selected.    

\begin{figure}[!htb]
\subfloat[Throughput of UE $u_1$]{
\includegraphics[scale=.95]{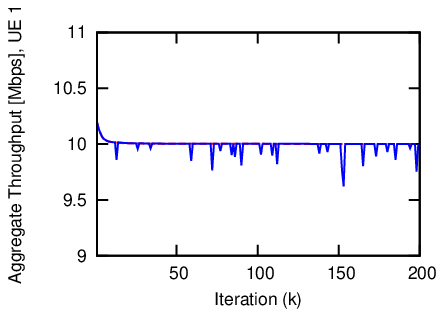}
\label{fig:THUE1}}
\subfloat[Throughput of UE $u_2$]{
\includegraphics[scale=.95]{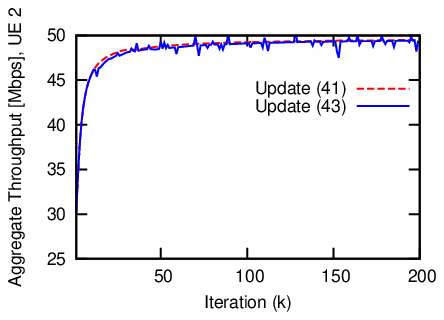}
\label{fig:THUE2}}\\
\subfloat[${z_{a,1}}$]{
\includegraphics[scale=.95]{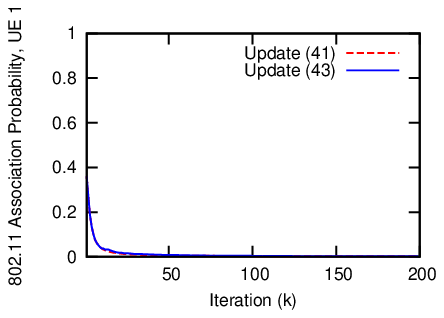}
\label{fig:ZUE1}}
\subfloat[${z_{a,1}}$]{
\includegraphics[scale=.95]{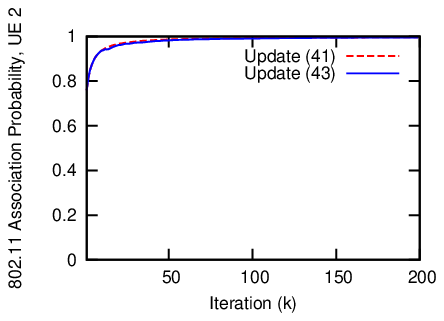}
\label{fig:ZUE2}}
\caption{Convergence of updates (\ref{eq:a1}) and (\ref{eq:a2}) when $\mathbf{X}_{k+1}$ is drawn uniformly random within the ball of radius $0.025$. 
}
\label{fig:ccpupdates}
\end{figure}

For comparison, Figure \ref{fig:ccpupdates} shows the convergence  of both updates (\ref{eq:a1}) and (\ref{eq:a2}).   Observe that in this example the extra complexity of update (\ref{eq:a2}) does not yield a better optimum, and indeed we also observe this in the other examples presented below (we also carried out further tests, not shown here, that exhibit similar behaviour).  This suggests that for the class of optimisation problems considered here update (\ref{eq:a1}) tends to converge to a near optimal solution.

\section{Example Scenarios}
\subsection{802.11 Offload}
Our problem formulation can be readily extended to provide a principled approach for offload of data trafiic from LTE to 802.11.  In an offload setting we would like to preferentially use the 802.11 network where possible.   We can capture this requirement by augmenting the proportional fair utility function $\sum_{u\in \U} \log \left(s_u + r_u\right)$ with a cost related to LTE usage.  For example, the cost function
\begin{align}
H(\zeta^i_{b,u})=\sum_{u \in \U} \sum_{b \in \B} \sum_{i \in \I} \frac{\zeta^i_{b,u}}{\beta_1 \omega^i \log{(1+\frac{\gamma^i_{b,u}}{\beta_2})}}\label{eq:H}
\end{align}
associates a cost with the LTE airtime used by each UE (the airtime being inversely proportional to the data rate).   

Consider a simple network setup with one BS, one AP and two UEs.   Suppose the 802.11 physical rate is 54Mbps for both users and the LTE rates are 4Mpbs (corresponding to an SNR of 0.75dB), see Table \ref{tb:wifioffloadrates}.  Other 802.11 LTE parameters are  detailed in Table \ref{tb:parameters}. 

\begin{figure}[!htb]
  \subfloat[t][]   
    {\includegraphics[scale=1]{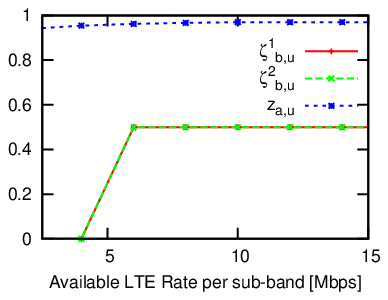}}
\subfloat[t][]
    {\includegraphics[scale=1]{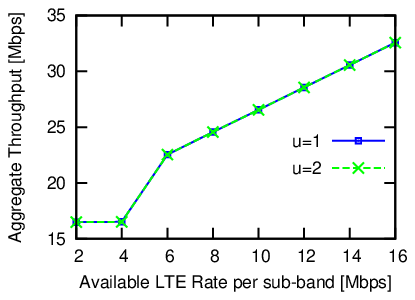}}
    \caption{Illustrating 802.11 offload as LTE data rate is varied, (a): LTE airtime, (b): Aggregate LTE plus 802.11 UE throughput. }
\label{fig:wifi_offload} 
\end{figure}

Using update (\ref{eq:a1}) to solve the optimisation problem, Table \ref{tb:wifioffloadresults} summarises the solution found.   The UEs share the 802.11 AP and do not send traffic via the LTE network (the LTE rates $r_{u_1}=0=r_{u_2}$).  The throughput of each UE is 16.5Mbps, which is higher than the data rate of 2Mbps provided by the LTE network. 
%
\begin{table}
\centering
\caption{802.11 offload example data rates.}
\label{tb:wifioffloadrates}
\begin{center}
\resizebox{\columnwidth}{!}{%
\begin{tabular}{|c|c|c|c|c|c|}
\hline
\multirow{2}{.1cm}{}& \multicolumn{2}{p{2cm}|}{\centering PHY Rates [Mbps]} & \multicolumn{3}{p{2.8cm}|}{\centering Technology Rates [Mbps]} \\
\cline{2-6} & \multicolumn{1}{c|}{BS}  & \multicolumn{1}{c|}{AP1}  & \multicolumn{1}{c|}{LTE only} & \multicolumn{1}{c|}{802.11 only}  & \multicolumn{1}{c|}{Optimised Multi-RAT} \\ \hline
$u_1$  & 4 & 54 & 2 & 16.5 & 16.5 \\
$u_2$ & 4 & 54 & 2 & 16.5 & 16.5 \\ 
\hline
\end{tabular}
}
\end{center}
\end{table}
%
%
%
\begin{table}
\caption{Rate allocation for 802.11 offload example.}
\label{tb:wifioffloadresults}
\centering
\begin{tabular}{|l|l|l|}
\hline
\multicolumn{1}{|l|}{Parameter}&\multicolumn{1}{l|}{$u_1$}&\multicolumn{1}{l|}{$u_2$}\\
\hline\hline
$r_u$ & $0$    &     $0$ \\
$\rho_{a,u}$ & $0.48$   &    $0.48$ \\
$s_u$ & $16.47$    &     $16.47$ \\
$p_{a,u,1}$ &   $0.046$    &    $ 0.046$ \\
$p_{a,u,2}$ & $0.954$     &    $0.954$ \\
$z_{a,u}$ & $0.95$  &      $0.95$ \\
\hline
\end{tabular}
\end{table}
Figure \ref{fig:wifi_offload}(a) plots how the fraction $\zeta^i_{1,b,u}$ of LTE airtime used by each UE changes as the LTE data rate is increased, while Figure \ref{fig:wifi_offload}(b) shows the aggregate throughput (LTE plus 802.11) of each UE.   It can be seen that for LTE data rates less than 5Mbps the LTE network is not used and data is fully offloaded to the 802.11 network.   However, as the LTE data rate increases the LTE network is increasingly 
 used to enhance the user throughputs.   Observe also from Figure \ref{fig:wifi_offload}(a) that the probability $z_{a,u}$ of associating to the 802.11 network remains almost constant as the use of the 802.11 network always results in an increase in the aggregate throughput. 
 
\begin{figure}[!htb]
  \subfloat[t][UE $u_1$]   
    {\includegraphics[scale=1]{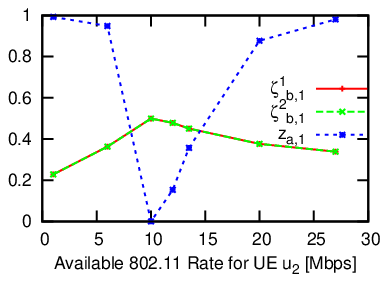}}
\subfloat[t][UE $u_2$]
    {\includegraphics[scale=1]{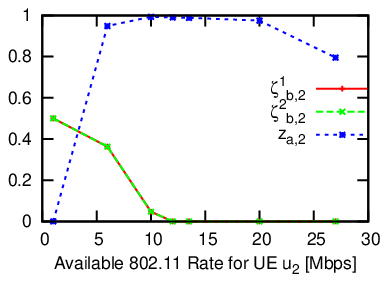}}
    \caption{Illustrating 802.11 offload as 802.11 rate is varied.  }
\label{fig:wifivarying_offload} 
\end{figure}

We can readily extend consideration to include situations where UEs have a specified traffic load by adding additional constraints to the optimisation problem.  For example, suppose that both UEs have a traffic load of 10Mbps video, the LTE available rate for each UE is 10Mbps (i.e. the total LTE capacity is 20 Mbps, enough to support the traffic load) and the 802.11 physical rate for UE $u_1$ is fixed at 6Mbps while that of UE $u_2$ is varied.  Using (\ref{eq:H}) to account for the cost of using the LTE connection we expect the traffic to be offloaded to the 802.11 network as long as its capacity is sufficient to meet the load.  Figure \ref{fig:wifivarying_offload} illustrates the LTE air time  for UEs $u_1$ and $u_2$ vs the 802.11 rate available to UE $u_2$. Since the 802.11 rate for $u_1$ is fixed at 6Mbps, $u_1$ must use the LTE connection in order to meet its traffc demand and indeed, as expected, it can be seen in Fig \ref{fig:wifivarying_offload}(a) that $\zeta_{b,1}^1$ and $\zeta_{b,1}^2$ (the fractions of LTE airtime used by UE $u_1$ on each subband) are always non-zero.  In contrast, it can be seen from Fig \ref{fig:wifivarying_offload}(b) that the LTE airtime used by UE $u_2$ falls to zero once the available 802.11 rate rises above 10Mbps \emph{i.e.} once sufficient 802.11 bandwidth is available to support the traffic load it is fully offloaded from LTE onto the 802.11 link.   Observe also that the 802.11 usage by UE $u_1$ falls to zero when the 802.11 rate is 10 Mbps.   This allows UE $u_2$ to make full use of the 802.11 network to meet its traffic demand, without incurring the overhead of collisions.   Once the 802.11 rate increases further, the extra capacity is then used by UE $u_1$.

\subsection{802.11 Multihoming}
We next illustrate how our framework can be used to manage 802.11 multihoming.   Consider an example with one BS,  two APs ($a_1$ and $a_2$) and two UEs ($u_1$ and $u_2$).  The available data rates are summarised in Table \ref{tb:wifimultirates}.    Each user can individually achieve 10Mbps over the LTE network (and 5Mbps each when sharing the LTE network).  The users have a physical rate of 36Mbps and 54Mbps over the first 802.11 WLAN and of 36Mbps and 18Mbps over the second WLAN, so user $u_2$ has a better connection via AP $a_1$ and user $u_1$ will have a better connection via AP $a_1$ by avoiding contention with user $u_2$.  Note that although 54Mbps link has a higher physical rate compared to that of the 36Mbps link, when the duration of a successful transmission is held constant the achievable throughputs are roughly equal due to the MAC framing and contention overhead.

Table \ref{tb:wmresults} details the rate allocation found using update (\ref{eq:a1}).   It can be seen that UE $u_1$ is mainly associated with AP $a_2$ ($z_{a_2,u_1}=0.99$) and $u_2$ with AP $a_1$ ($z_{a_1,u_2}=0.99$), as might be expected in view of the link characteristics.    The LTE network is shared equally by both users ($\zeta_{b,u_1}=0.5=\zeta_{b,u_2}$).   
\begin{table}
\centering
\caption{802.11 multihoming example data rates.}
\label{tb:wifimultirates}
\begin{center}
\resizebox{\columnwidth}{!}{%
\begin{tabular}{|c|c|c|c|c|c|c|c|}
\hline
\multirow{2}{.1cm}{}& \multicolumn{3}{p{2cm}|}{\centering PHY Rates [Mbps]} & \multicolumn{4}{p{2.8cm}|}{\centering Technology Rates [Mbps]} \\
\cline{2-8} & \multicolumn{1}{c|}{BS}  & \multicolumn{1}{c|}{AP $a_1$} & \multicolumn{1}{c|}{AP $a_2$} & \multicolumn{1}{c|}{LTE} & \multicolumn{1}{c|}{WLAN $a_1$} & \multicolumn{1}{c|}{WLAN $a_2$} & \multicolumn{1}{c|}{Optimised Multi-RAT} \\ \hline
$u_1$  & 10 & 36 & 36 & 5 & 5.9 & 8.9 & 22.39 \\
$u_2$  & 10 & 54 & 18 & 5 &8.87 & 4.3 & 22.39 \\ 
\hline
\end{tabular}
}
\end{center}
\end{table}
%
%

\begin{table}
\caption{Rate allocation for 802.11 multihoming example.}
\label{tb:wmresults}
\centering
\begin{tabular}{|l|l|l|}
\hline
\multicolumn{1}{|l|}{Parameter}&\multicolumn{1}{l|}{$u_1$}&\multicolumn{1}{l|}{$u_2$}\\
\hline\hline
$\zeta_{b,u}$ & $ 0.5$  &    $0.5$ \\
$\rho_{a_1,u}$ & $0.0045$   &    $0.92$ \\
$\rho_{a_2,u}$ & $0.94$   &    $0.0037$ \\
$r_u$ & $5$ & $5$ \\
$s_u$  & $17.39$    &     $17.4$ \\
$p_{a_1,u,1}$ &   $0.007$    &    $ 1$ \\
$p_{a_2,u,1}$ & $1$ &   $0.007$ \\
$z_{a_1,u}$ & $0$  &      $0.99$ \\
$z_{a_2,u}$ & $0.99$  &      $0$ \\
\hline
\end{tabular}
\end{table}
%

\begin{figure}[!htb]
  \subfloat[t][UE $u_1$]   
    {\includegraphics[scale=1]{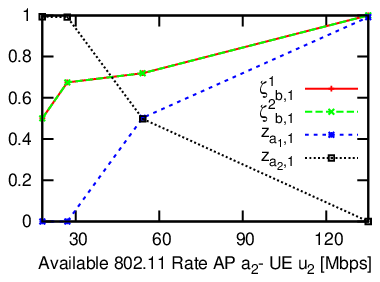}}
\subfloat[t][UE $u_2$]
    {\includegraphics[scale=1]{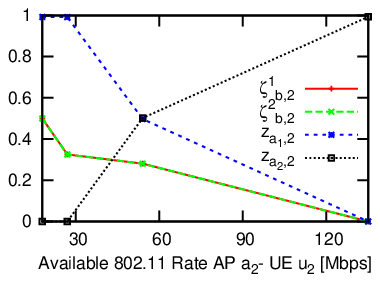}}
    \caption{Illustrating 802.11 multi-homing as the rate on the UE $u_2$- AP $a_2$ 802.11 link is varied.}
\label{fig:wifimulti_varying} 
\end{figure}

Figure \ref{fig:wifimulti_varying} shows how the allocation changes as the capacity of the 802.11 link between AP $a_2$ and UE $u_2$ varies.  It can be seen that as the rate on the $a_2$-$u_2$ link increases UE $u_2$ increasingly makes use of this link ($z_{{a_2},2}$ increases while $z_{{a_1},2}$ falls) and conversely traffic for UE $u_1$ increasingly makes use of AP $a_1$ ($z_{{a_1},1}$ increases while $z_{{a_2},1}$ falls) and of the LTE network ($\zeta_{b,1}^1$ and $\zeta_{b,1}^2$ both rise).   When both UEs have equal rates of 36Mbps and 54Mbps through APs $a_1$ and $a_2$ respectively, they make use of APs equally 
%
\subsection{LTE Multihoming}
We next consider an example with two BSs ($b_1$ and $b_2$), one AP, and 4 UEs ($u_1$, $u_2$, $u_3$, and $u_4$). The physical rates available on the 802.11 and LTE links are summarised in Table \ref{tb:ltemultirates}.  We also consider the conventional user assignments to LTE BSs \emph{i.e.} each UE associates with the LTE BS that provides the UE with the maximum received power. The achievable LTE rates when using this maximum received power assignment are  also summarised in  Table \ref{tb:ltemultirates}. 
%
%
\begin{table}[!htb]
\centering
\caption{LTE multihoming example: data rates.}
\label{tb:ltemultirates}
\begin{center}
\resizebox{\columnwidth}{!}{%
\begin{tabular}{|c|c|c|c|c|c|c|c|c|}
\hline
\multirow{2}{.1cm}{}& \multicolumn{3}{p{2cm}|}{\centering PHY Rates [Mbps]} & \multicolumn{5}{p{2.8cm}|}{\centering Technology Rates [Mbps]} \\
\cline{2-9} & \multicolumn{1}{c|}{BS $b_1$}  & \multicolumn{1}{c|}{BS $b_2$} & \multicolumn{1}{c|}{AP} & \multicolumn{1}{c|}{LTE $b_1$ only} & \multicolumn{1}{c|}{LTE $b_2$ only} & \multicolumn{1}{c|}{LTE (Maximum Rx Power)} &\multicolumn{1}{c|}{802.11 only} & \multicolumn{1}{c|}{Optimised Multi-RAT} \\ \hline
$u_1$  & 26 & 25 & 54 & 6.5 & 6.25 & 8.27 &7.51 & 15.36  \\
$u_2$  & 10 & 25 & 27 & 2.5 & 6.25 & 2.4 & 3.75 & 12.5  \\ 
$u_3$  & 5 & 29 & 54 & 1.25 & 7.25 &  3.56 &7.51 & 14.8  \\
$u_4$  & 11 & 10 & 13.5 & 2.75 & 2.5 & 2.29 &1.88 & 5.5 \\ 
\hline
\end{tabular}
}
\end{center}
\end{table}

\begin{figure}
    \includegraphics[scale=1]{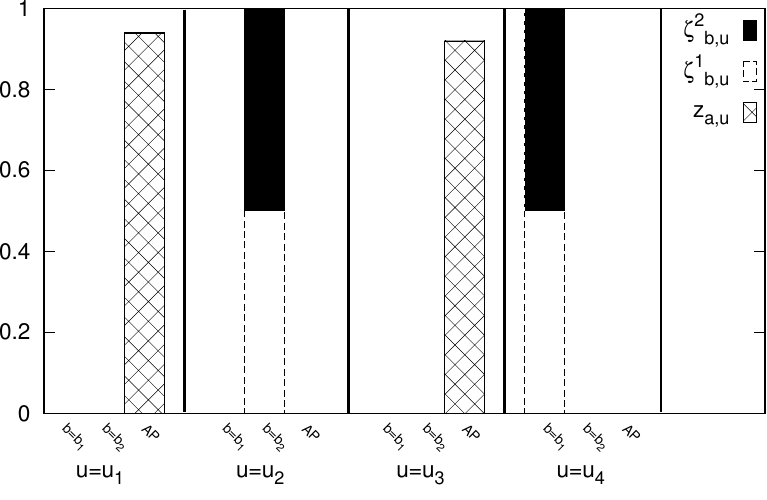}
\caption{ LTE multihoming, user association probabilities per sub-band of the LTE BS ($\zeta^i_{b,u}$) and per 802.11 AP ($z_{a,u}$) given by Update (\ref{eq:a1}). }
\label{fig:apltemulti}
\end{figure}

Figure \ref{fig:apltemulti} shows the proportional fair LTE sub-band allocations and 802.11 association probabilities obtained using update (\ref{eq:a1}).   From Table \ref{tb:ltemultirates} it can be seen that a relatively large physical rate of 54 Mbps is available to UEs $u_1$ and $u_3$ and so it can be seen from Figure \ref{fig:apltemulti} that in the proportional fair allocation these UEs make use of the 802.11 network but not the LTE network, so freeing up capacity in the LTE network for UEs $u_2$ and $u_4$.  UEs $u_2$ and $u_4$ make use of the LTE rather than the 802.11 network (so reducing collisions and increasing the 802.11 capacity available to UEs $u_1$ and $u_3$), and share the LTE network evenly between them.   

It can be seen from the right-hand column of Table \ref{tb:ltemultirates} that all users benefit from this use of multi-homing.  In addition  observe that a simple aggregation of both LTE (Maximum Power association) and 802.11 resources results in 13\% reduction in the proportional fair rate objective compared with that of a near-optimal solution.



%

\section{Summary and Conclusion}
We consider proportional fair rate allocation in a heterogeneous network with a mix of LTE and 802.11 cells which supports multipath and multihomed operation (simultaneous connection of a user device to multiple LTE BSs and 802.11 APs).  We show that the utility fair optimisation problem is non-convex but that a global optimum can be found by solving a sequence of convex optimisations in a distributed fashion.  The result is a principled approach to offload from LTE to 802.11 and for exploiting LTE/802.11 path diversity to meet user traffic demands.

\section*{Appendix A} \label{sec:appdx1}
\subsection{Optimisation problem}
With the obvious abuse of notation we add subscripts to the functions $h^{(i)}$ and $g^{(i)}$ to streamline the presentation since there is no risk of confusion.  Details of the objective and constraints functions are as follows:
\begin{align*}
&f(\mathbf{x})=- \sum_{u \in \U} \log(s_u+r_u)\\
%
&h^{(1)}_{u}(\mathbf{x})=s_u\\
&h^{(2)}_{a,u}(\mathbf{x})=\tilde{\rho}_{a,u}-\tilde{w}_{a,u}+\sum_{v\in \U_a} \log(1+e^{\tilde{w}_{a,v}}) \nonumber \\
&\qquad\qquad - \log \left(\sum_{n=1}^{|\U_a|} \frac{T_{s_{a}}}{T_c}\frac{\phi}{\Phi_n}q_{a,u,n} \right)\\
&h^{(3)}_{a,u,n}(\mathbf{x})=q_{a,u,n}, 
~h^{(4)}_{u}(\mathbf{x})=\sum_{a \in \A_u} e^{\tilde{w}_{a,u}} \\
&h^{(5)}_{i}(\mathbf{x})=\sum_{u \in \U} \sum_{b\in \B}\zeta^i_{b,u}-1\\
& h^{(6)}_{u}(\mathbf{x})=r_u-\sum_{b \in \B} \sum_{i \in \I} \zeta^{i}_{b,u} \beta_1 \omega^i \log{(1+\frac{\gamma^{i}_{b,u}}{\beta_2})} 
\end{align*}

\begin{align*}
&g^{(1)}_{u} (\mathbf{x})= \sum_{a \in \A_u} e^{\tilde{\rho}_{a,u}} c_{a,u},  ~g^{(2)}_{a,u}(\mathbf{x})=0 \\
&g^{(3)}_{a,u,n}(\mathbf{x})=\sum\limits_{\stackrel{\tilde{\U}_a\in}{\mathcal{P}_{n-1}(\U_a \setminus \{u\} )}}  \prod_{v\in \tilde{\U}_a}e^{\tilde{w}_{a,v}}\\
&g^{(4)}_{u}(\mathbf{x})=\sum_{a \in A_u} \frac{e^{2\tilde{w}_{a,u}}}{1+e^{\tilde{w}_{a,u}}},
~g^{(5)}_i(\mathbf{x})=0, ~g^{(6)}_u(\mathbf{x})=0
\end{align*}
\normalsize
\section*{Appendix B} \label{sec:appdx2}
\subsection{Proof of Theorem \ref{lem:one}}
\begin{proof}
Update (\ref{eq:a1}) generates a sequence $\{\mathbf{x}_k, k=1,2,\cdots\}$.  By the convexity of $g^{(i)}$ we have that that $g^{(i)}(\mathbf{x}) \ge g^{(i)}(\mathbf{x}_k) + \partial g^{(i)}(\mathbf{x}_k)(\mathbf{x}-\mathbf{x}_k)$ and so $C_{\mathbf{x}_k}\subset C$ (for any $\mathbf{x}\in C_{\mathbf{x}_k}$ we have $h^{(i)}(\mathbf{x}) - g^{(i)}(\mathbf{x}) \le h^{(i)}(\mathbf{x}) - g^{(i)}(\mathbf{x}_k) - \partial g^{(i)}(\mathbf{x}_k)(\mathbf{x}-\mathbf{x}_k) \le  0$, $i=1,\cdots,m$).  Hence, $\mathbf{x}_k\in C$, $k=1,2,\cdots$.   Further, $f(\mathbf{x}_{k+1})\le f(\mathbf{x}_k)$ and so the sequence $\{f(\mathbf{x}_k)\}$ is decreasing.  Since $f$ is convex (so continuous) and $C\subset B$ is bounded then $f(\mathbf{x}_k)$ is bounded.  Hence, by the monotone convergence of bounded sequences (\emph{e.g.} \cite[Theorem 16.2]{billingsley2008probability}), sequence $\{f(\mathbf{x}_k)\}$ converges to a finite limit.  Let $f_\infty$ denote this limit and let $C_\infty=\{\mathbf{x} \in C: f(\mathbf{x})=f_\infty\}$ denote the corresponding set of limit points in $C$.    

From (\ref{eq:a1}) we have that for every $\mathbf{x}\in C_\infty$ then $\mathbf{x}\in \arg \min_{\mathbf{x}\in C_{\mathbf{x}} f(\mathbf{x})}$ (else we could find a point $\mathbf{y}\in D_{\mathbf{x}}$ such that $f(\mathbf{y})<f(\mathbf{x})$ contradicting the fact that $\mathbf{x}$ corresponds to a limit point of monotonic sequence $\{f(\mathbf{x}_k)\}$).   It follows that $\mathbf{x}$ satisfies the Fritz John conditions,
\begin{align}
K_0(\mathbf{x}) &:= \lambda_0\partial f(\mathbf{x}) + \sum_{i=1}^m \lambda^{(i)}(\partial h^{(i)}(\mathbf{x}) - \partial g^{(i)}(\mathbf{x})) = 0\\
K_i(\mathbf{x}) &:= \lambda^{(i)}(h^{(i)}(\mathbf{x})-g^{(i)}(\mathbf{x})) = 0,\ i=1,\cdots,l
\end{align}
 with multipliers $\lambda^{(i)}\ge 0$, $i=0,\cdots,l$.   But these are also the Fritz John conditions for optimisation P, and so it follows that $\mathbf{x}$ is a stationary point of P.   Since $f$ is convex it is Lipschitz continuous on compact set $C$ and so $f(\mathbf{x}_k)\rightarrow f_\infty$ implies that for any $\delta>0$ and $k\ge k_\delta$, $\exists \mathbf{x}\in C_\infty$ s.t. $\|\mathbf{x}_k-\mathbf{x}\|<\delta$  provided $k_\delta$ is sufficiently large.  By assumption, $\partial f$, $\partial h^{(i)}$ and $\partial g^{(i)}$ are continuous, so $K_0$, $K_i$ are continuous.  Hence, $K_0(\mathbf{x}_k)\rightarrow 0$, $K_i(\mathbf{x}_k)\rightarrow 0$, $i=1,\cdots,l$ and we are done.
 \end{proof}
\subsection{Proof of Theorem \ref{lem:two}}
\begin{proof}
Since $C$ is compact, then for every open covering there exists a finite subcovering.  Let $\cup_{\Upsilon\in \Psi \subset 2^C} \Upsilon$ be a covering of $C$ consisting of balls of radius $r$ and let $\cup_{\Upsilon \in \Psi^\prime\subset \Psi} \Upsilon$ be a finite subcovering.   Each set $\Upsilon \in \Psi^\prime$ either covers $C$ or has at least one neighbour $\Sigma \in \Psi^\prime$ such that $\Upsilon \cap \Sigma \ne \emptyset$ since $C$ is connected.   Now expand each set $\Upsilon \in \Psi^\prime$ to be a ball of radius $2r$ and define a new set $\tilde{\Upsilon}$ obtained by taking the union of the expanded set $\Upsilon$ with all of its expanded neighbours.   In the new expanded covering $\tilde{\Psi}^\prime$ so obtained neighbours $\tilde{\Upsilon}$ and $\tilde{\Sigma}$ have at least a ball of radius $r$ in common, so $\tilde{\Upsilon}\cap\tilde{\Sigma}$ has volume at least $\nu(r)$ where $\nu(r)$ is the volume of a hypersphere of radius $r$ in $\mathbb{R}^n$.  Since $C$ is connected and the sets in $\tilde{\Psi}^\prime$ form a finite covering, between any two points $\mathbf{x}$, $\mathbf{y}\in C$ there exists a path traversing a sequence of at most $|\tilde{\Psi}^\prime|$ neighbouring sets from $\tilde{\Psi}^\prime$.   Given $\mathbf{x}_k\in \tilde{\Upsilon}$ then the probability that $\mathbf{x}_{k+1}$ lies in a neighbour $\tilde{\Sigma}$ is at least $\epsilon\eta\nu(r)$ and so every path is traversed with probability at least $(\epsilon\eta\nu(r))^{|\tilde{\Psi}^\prime|}$.   That is, starting from any initial condition for every optimum $\mathbf{x}^*$ of problem P a set $W$ of size no greater than $\nu(2r)$ containing $\mathbf{x}^*$ is visited with probability at least $(\epsilon\eta\nu(r))^{|\tilde{\Psi}^\prime|}$ .   This holds for every $r>0$.  Selecting $\xi = \nu(2r)$, then for any initial condition and every $\xi>0$, for every optimum $x^*$ of optimisation $P$ a ball $B_\xi(x^*)$ containing $x^*$ is visited with positive probability and the stated result now follows.
\end{proof}
\bibliography{references}
\bibliographystyle{ieeetr}
\begin{biography}{Bahar Partov} is pursuing a PhD degree at the Hamilton Institute together with Bell-labs Alcatel-Lucent Ireland. She received her master's degree from University of Essex at 2009. She did her undergraduate degree at University of Tabriz, Iran. Her current research interests are distributed algorithms for self-organized networks.
\end{biography}
\begin{biography}{Doug Leith} graduated from the University of Glasgow in 1986 and was awarded his PhD, also from the University of Glasgow, in 1989. In 2001, Prof. Leith moved to the National University of Ireland, Maynooth and then in Dec 2014 to Trinity College Dublin to take up the Chair of Computer Systems in the School of Computer Science and Statistics.  His current research interests include wireless networks, network congestion control, distributed optimisation and data privacy.   
\end{biography}

\end{document}